\documentclass[twocolumn,9pt]{extarticle}
\usepackage[top=1.8cm, bottom=2cm, left=1.75cm, right=1.75cm]{geometry}

\usepackage[T1]{fontenc}
\usepackage{baskervillef}
\usepackage{flushend}
\usepackage[varqu,varl,var0]{inconsolata}
\usepackage[scale=.95,type1]{cabin}
\usepackage[baskerville,vvarbb]{newtxmath}
\usepackage[cal=boondoxo]{mathalfa}

\usepackage[hyphens]{url}
\usepackage{graphicx}  %
\usepackage[hyphens]{url}  %
\usepackage[hang,flushmargin]{footmisc}
\renewcommand{\footnotesize}{\fontsize{8}{9}\selectfont}

\makeatletter
\def\url@leostyle{%
  \@ifundefined{selectfont}{\def\UrlFont{}}%
  {\def\UrlFont{}}%
}
\makeatother
\usepackage[numbers,sort&compress]{natbib}
\usepackage{color}

\usepackage{xspace}
\newcommand{\uniform}{{\tt uniform}\xspace}
\newcommand{\quantile}{{\tt quantile}\xspace}
\newcommand{\kmeans}{{\tt k-means}\xspace}
\newcommand{\privtree}{{\tt PrivTree}\xspace}

\usepackage[T1]{fontenc}

\newcommand{\reduce}{\vspace{-0.0cm}}

\usepackage{amsmath}
\usepackage{booktabs}

\usepackage{paralist}
\usepackage{enumitem}
\setlist[itemize]{itemsep=0pt}
\setlist[enumerate]{itemsep=0pt}

\usepackage{graphicx}	%
	\setkeys{Gin}{width=\textwidth, height=\textheight, keepaspectratio}

\usepackage[bf]{caption}

\usepackage[skip=3pt,justification=centering]{subcaption}
\newif\ifcomment
\commenttrue
\commentfalse
\ifcomment
	\newcommand{\edc}[1]{\textbf{\em\color{red}EDC: #1}}
	\newcommand{\gvg}[1]{\textbf{\em\color{blue}GG: #1}}
	\newcommand{\sundar}[1]{\textbf{\em\color{purple}MS: #1}}
	\newcommand{\sof}[1]{\textbf{\em\color{brown}SOF: #1}}
    \newcommand{\chg}[1]{{\color{change}#1}}
    \newcommand{\chgTag}[2]{{\color{change}{\bf [#1]} #2}}
\else
	\newcommand\edc[1]{}
	\newcommand\gvg[1]{}
	\newcommand\sundar[1]{}
	\newcommand\sof[1]{}
    \newcommand{\chg}[1]{#1}
    \newcommand{\chgTag}[2]{#2}
\fi

\newcommand{\descr}[1]{\smallskip\noindent\textbf{#1}}

\usepackage{cancel}

\newcommand{\mywidth}{0.99}

\newcommand{\sscale}{0.95}

\usepackage[bookmarks=true, bookmarksnumbered=true, colorlinks=true, linkcolor=linkcol, citecolor=citecol, urlcolor=urlcol, hypertexnames=true]{hyperref}

\definecolor{darkgreen}{RGB}{0, 100, 0}
\definecolor{linkcol}{rgb}{0.3,0,0}
\definecolor{citecol}{rgb}{0.3,0,0}
\definecolor{urlcol}{rgb}{0.3,0,0}
\definecolor{vlightgray}{gray}{0.925}

\let\OLDthebibliography\thebibliography
\renewcommand\thebibliography[1]{
  \OLDthebibliography{#1}
  \setlength{\parskip}{0pt}
  \setlength{\itemsep}{1pt plus 0.2ex}
}

\usepackage[capitalise,noabbrev]{cleveref}
\newcommand{\plainfootnote}[1]{%
  \let\thefootnote\relax%
  \footnotetext{#1}%
}

\begin{document}

\sloppy
\title{\huge \bf The Importance of Being Discrete: Measuring the Impact of Discretization in End-to-End Differentially Private Synthetic Data$\mathbf{^*}$}
\date{}

\author{Georgi Ganev$^{1,2}$, Meenatchi Sundaram Muthu Selva Annamalai$^{1}$,\\Sofiane Mahiou$^{2}$, and Emiliano De Cristofaro$^{3}$\\[1ex]
$^1$UCL, $^2$SAS, $^3$UC Riverside}

\maketitle

\let\origthefootnote\thefootnote
\let\thefootnote\relax\footnotetext{$\mathbf{^*}$Published in the Proceedings of the 32nd ACM Conference on Computer and Communications Security (ACM CCS 2025).}

\begin{abstract}
Differentially Private (DP) generative marginal models are often used in the wild to release synthetic tabular datasets in lieu of sensitive data while providing formal privacy guarantees.
These models approximate low-dimensional marginals or query workloads; crucially, they require the training data to be pre-discretized, i.e., continuous values need to first be partitioned into bins.
However, as the range of values (or their domain) is often inferred directly from the training data, with the number of bins and bin edges typically defined arbitrarily, this approach can ultimately break end-to-end DP guarantees and may not always yield optimal utility.

In this paper, we present an extensive measurement study of four discretization strategies in the context of DP marginal generative models.
More precisely, we design DP versions of three discretizers (uniform, quantile, and k-means) and reimplement the PrivTree algorithm.
We find that optimizing both the choice of discretizer and bin count can improve utility, on average, by almost 30\% across six DP marginal models, compared to the default strategy and number of bins, with PrivTree being the best-performing discretizer in the majority of cases.
We demonstrate that, while DP generative models with non-private discretization remain vulnerable to membership inference attacks, applying DP during discretization effectively mitigates this risk.
Finally, we \chgTag{C9}{improve on an existing} approach for automatically selecting the optimal number of bins, and achieve high utility while reducing both privacy budget consumption and computational overhead.

\end{abstract}
\smallskip
\smallskip

\let\thefootnote\origthefootnote

\section{Introduction}
\label{sec:intro}
Privacy-preserving synthetic data promises to enable the sharing of sensitive data %
by training a generative model to learn its underlying distribution while limiting individual-level information leakage%
~\cite{rs2023privacy, un2023guide, oecd2023emerging}.
More precisely, leakage can be bound by satisfying Differential Privacy (DP)~\cite{dwork2006calibrating, dwork2014algorithmic} via randomized mechanisms that add calibrated noise to the learning procedure.
The trained model can then be sampled to generate and release an arbitrary number of synthetic datasets without any further privacy leakage.

In the context of tabular data, DP synthetic data \chg{ostensibly} offers several benefits: high utility on downstream tasks, \chg{seamless} integration into existing data pipelines, ability to be safely regenerated/reused, etc.~\cite{jordon2022synthetic, rs2023privacy, fca2024using}.
DP synthetic data is being adopted beyond academic research~\cite{jordon2022synthetic, cristofaro2024synthetic, hu2024sok}, by startups~\cite{forbes2022synthetic, techcrunch2022the}, corporations~\cite{microsoft2022iom, ibm2024creating, sas2024data}, government agents~\cite{nist2018differential, nist2020differential, nasem2020census, ons2023synthesising}, regulators~\cite{ico2023privacy, ico2023synthetic, fca2024using}, and nonprofits~\cite{rs2023privacy, un2023guide, oecd2023emerging}.

While there is no single universally ``best'' model~\cite{jordon2022synthetic}, benchmark studies~\cite{tao2021benchmarking, ganev2024graphical}, \chgTag{C1}{real-world deployments~\cite{ons2023synthesising, ico2023synthetic, hod2025differentially}, and NIST competitions~\cite{nist2018differential, nist2020differential}} have consistently ranked marginal models (including graphical~\cite{zhang2017privbayes, mckenna2021winning} and workload/query-based approaches~\cite{aydore2021differentially, liu2021iterative, mckenna2022aim}) as providing the best privacy-utility tradeoffs compared to deep-learning based techniques. %
Marginal models are particularly effective for relatively small tabular datasets (fewer than 32 features) and simple tasks, e.g., capturing summary statistics and replicating them in the synthetic data~\cite{tao2021benchmarking, ganev2024graphical}.
Broadly, they follow the {\em select–measure–generate} paradigm~\cite{mckenna2021winning, mckenna2022simple}: i) select a collection of marginals of interest, ii) measure and add noise to the marginals, and iii) generate synthetic data that preserves them.

\begin{figure*}[t!]
\begin{minipage}[t]{0.49\textwidth}
	\includegraphics[width=\linewidth]{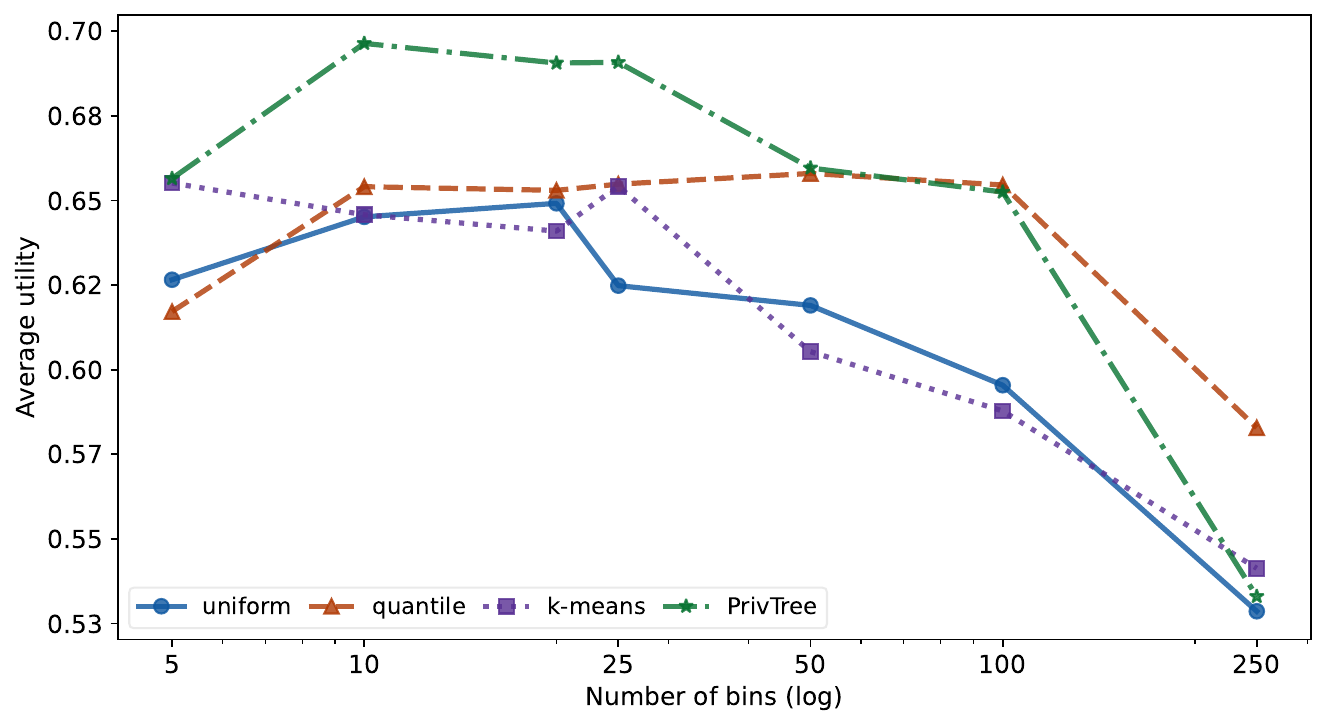}
  \caption{Utility of four DP discretizers, averaged across six DP generative models ($\epsilon=1$) and three datasets (US3).}
	\label{fig:disc_comp_exp3}
  \end{minipage}
\hfill
\begin{minipage}[t]{0.49\textwidth}
	\includegraphics[width=\linewidth]{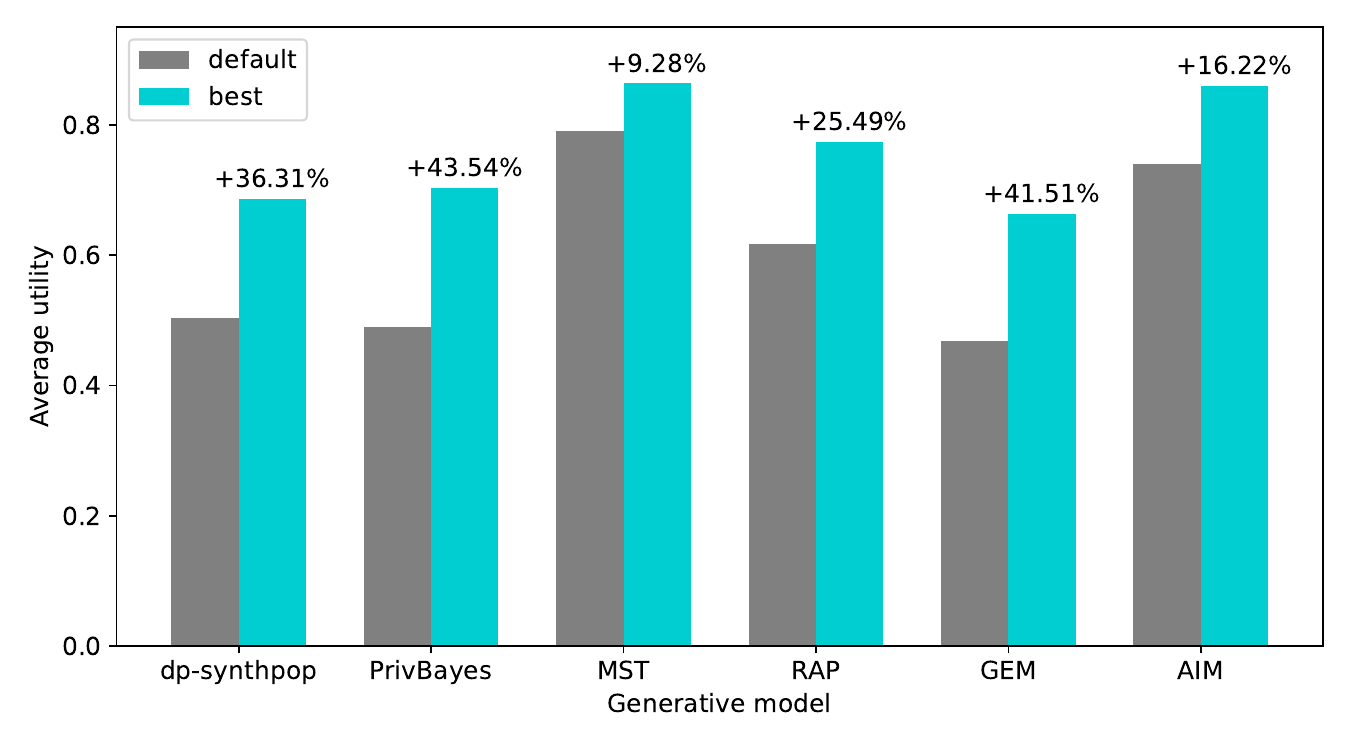}
  \caption{Utility of default discretizer (\uniform, 20 bins) and optimal discretizer/bin count for six DP generative models ($\epsilon=1$), averaged across three datasets (US3).}
	\label{fig:delta}
  \end{minipage}
\end{figure*}

\descr{Discretization.}
Marginal models typically require the input data to be pre-processed into discrete numerical values~\cite{zhang2017privbayes, aydore2021differentially, liu2021iterative, mckenna2022aim,mckenna2021winning}
to facilitate integrating efficient DP mechanisms~\cite{mckenna2022simple}.
This limits their applicability to real-world tabular datasets, as these typically consist of a mix of continuous and discrete columns.
As a result, it is common to discretize the data manually before feeding it as input to DP synthetic data algorithms~\cite{zhang2017privbayes, ping2017datasynthesizer, vietri2020new, mckenna2022aim, mahiou2022dpart, qian2023synthcity, du2024towards,mckenna2021winning}.

Discretization involves two steps: determining the domain of the raw data and assigning the raw values to bins.
In the first step, the minimum and maximum values for numerical columns and the list of possible categories for categorical columns are extracted.
In theory, this could be done using publicly available sources (e.g., data dictionaries or auxiliary datasets); in practice, however, the values are usually extracted directly from the input data~\cite{zhang2017privbayes, ping2017datasynthesizer, vietri2020new, mckenna2021winning, mckenna2022aim, mahiou2022dpart, qian2023synthcity, du2024towards}.
In the second step, continuous values are partitioned into bins and mapped to the corresponding indices.
While there are different ways to define the bins, for DP synthetic tabular data, the standard strategy in prior work is to use uniform discretization (i.e., equal-width bins), with the number of bins typically chosen arbitrarily between 8 and 50~\cite{zhang2017privbayes, ping2017datasynthesizer, vietri2020new, mckenna2021winning, mckenna2022aim, mahiou2022dpart, qian2023synthcity, du2024towards}.
Categorical values, on the other hand, are mapped using ordinal encoding.

\descr{Problem Statement.}
Despite the prevalence of discretization in DP generative marginal models, it remains unclear
\emph{how} and \emph{to what extent} this step affects privacy-utility tradeoffs.
For instance, the standard uniform discretization approach may not capture the underlying distribution accurately or be compatible with specific generative models, thus reducing the potential utility of the DP synthetic data.
Furthermore, it remains unclear how the number of bins, a common hyperparameter across discretization strategies, may affect utility and, more broadly, whether an optimal choice of bins exists across different discretization strategies, generative models, and privacy budgets.
Finally, and perhaps %
more importantly, the discretization step can by itself compromise the end-to-end DP pipeline if information is inferred directly from the input data, introducing unaccounted privacy leakage and %
violating DP~\cite{stadler2022synthetic, annamalai2024what}.

\descr{Research Questions.}
In this paper, we shed light on the impact of discretization on DP synthetic data's privacy-utility tradeoffs.
More precisely, %
we aim to answer the following research questions:

\begin{itemize}[leftmargin=2em]
	\item[$\bullet$] RQ1: How does discretization affect how well generative models capture and preserve controlled distributions?\smallskip
	\item[$\bullet$] RQ2: What is the impact of discretization on state-of-the-art DP marginal models, and is there a discretization strategy that performs best?\smallskip
	\item[$\bullet$] RQ3: What is the best method to determine the optimal number of bins for discretization strategies?\smallskip
	\item[$\bullet$] RQ4: How does domain extraction affect the privacy-utility tradeoff of DP marginal models?
\end{itemize}

\descr{Technical Roadmap.}
To answer these questions, we perform an extensive measurement study of the impact of discretization on synthetic data generation.
Our very first step is to integrate four discretization strategies into an end-to-end DP generative model pipeline.
More precisely, we convert three popular discretizers (\uniform, \quantile, and \kmeans) to make them differentially private %
and reimplement the already-DP \privtree algorithm~\cite{zhang2016privtree}, as we could not find a publicly available implementation.
In the process, we also rely on~\citet{desfontaines2020lowering}'s method to estimate the domain of numerical data (i.e., the minimum and maximum values) from the training data while satisfying DP.

Next, we measure the utility of the four discretization strategies on five controlled distributions with varying modality and skewness.
We then assess their performance vis-\`a-vis five state-of-the-art marginal generative models (PrivBayes~\cite{zhang2017privbayes}, MST~\cite{mckenna2021winning}, RAP~\cite{aydore2021differentially}, GEM~\cite{liu2021iterative}, and AIM~\cite{mckenna2022aim}), as well as a simple baseline model, dp-synthpop~\cite{mahiou2022dpart}, on three popular datasets (Adult, Gas, and Wine).
We experiment with a wide range of bin counts and evaluate four existing methods as well as an optimized approach for automatically selecting the number of bins to use.
We also show that the latter yields the best utility.

Finally, we examine the impact of applying DP to data domain extraction with different discretizers on the privacy leakage from the synthetic data using \chgTag{C7}{two} well-known inference attack\chg{s}.
Throughout our experiments, we evaluate the discretizers and generative models under varying privacy budgets --
in total, we fit over 300,000 discretizers and train around 200,000 generative models.

\descr{Main Findings.}
Our study yields a few interesting findings:
\begin{enumerate}[leftmargin=2em]
  \item When DP is only applied to the discretization process, increasing the number of bins consistently results in distributions being captured more accurately (\textit{RQ1}).
  \item Marginal models are highly sensitive to the discretization strategy and number of bins used.
	Their utility follows an inverted u-shaped trend as the number of bins increases: it initially improves but then degrades when too many bins introduce too much noise, as illustrated in~\cref{fig:disc_comp_exp3} (\textit{RQ2}).
  \item Optimizing both the discretizer and the number of bins can significantly help utility, with an improvement ranging from 9.28\% to 43.54\% compared to the default discretization (\uniform with 20 bins), as shown in~\cref{fig:delta} (\textit{RQ2}).
  \item Across our experiments, \privtree~\cite{zhang2016privtree} consistently delivers the best average performance, demonstrates minimal variability, and is computationally efficient (\textit{RQ2}).
  \item Our optimized method for automatic bin count selection outperforms alternatives and provides a 41\% normalized utility improvement \chgTag{C8}{relative to the range between the default and the best achievable utility via exhaustive search} (\textit{RQ3}).
  \item Although extracting the data domain with DP slightly reduces utility ($\approx$4\% on average), it substantially reduces inference attacks' success from 100\% (perfect inference) to $\approx$50\% (near-random guessing) compared to non-private extraction (\textit{RQ4}).
\end{enumerate}

Overall, our study demonstrates the importance of discretization in end-to-end DP pipelines for synthetic tabular data generation, an essential step under-explored and un-optimized by prior work.
By doing so, we not only show that substantial improvements in synthetic data utility can be made but also that discretization itself needs to be done in a DP manner, or these pipelines could be left exposed to serious privacy vulnerabilities.
We are confident that our work will help developers and practitioners be more conscious of the implementation details surrounding DP synthetic data generation and result in models with better privacy-utility tradeoffs.

\section{Background}
\label{sec:prelim}
This section presents background information on differential privacy, synthetic data generation, discretization, and privacy attacks.

\descr{Differential Privacy (DP)%
~\cite{dwork2006calibrating, dwork2014algorithmic}.}
A randomized algorithm $\mathcal{A}$ satisfies $(\epsilon, \delta)$-DP if, for all $S \subseteq \text{Range}(\mathcal{A})$ and all neighboring datasets $D$ and $D^{\prime}$ (differing in a single record), it holds that:
\begin{equation*}
\Pr[{\mathcal{A}}(D)\in S]\leq \exp \left(\epsilon \right)\cdot \Pr[{\mathcal{A}}(D^{\prime})\in S] + \delta
\end{equation*}

In other words, DP guarantees that, by examining the algorithm's output, an adversary cannot infer (up to the privacy budget, $\epsilon$) whether or not the data of a specific individual was part of the input dataset.
The parameter $\delta$ represents a small probability of failure to provide this guarantee.

Typically, DP is satisfied by applying randomized mechanisms that introduce noise to various steps of the algorithm.
DP also has two desirable properties, composition and post-processing, which, respectively, enable tracking the overall privacy budget when multiple mechanisms are used and allow for the reuse of DP-trained models without incurring additional privacy risks.

\descr{Discretization.}
The process of mapping a continuous value $\underline{x} \leq x \leq \overline{x}$ into a discrete value $\hat{x} \in \{1, 2, \dots, b\}$ is known as discretization.
This involves partitioning the domain of $x$ ($[\underline{x}, \overline{x}]$) into $b$ non-overlapping intervals, or bins, defined by their edges $\{e_1, e_2, \dots, e_{b+1}\}$, where $e_1 = \underline{x}$ and $e_{b+1} = \overline{x}$ denote the minimum and maximum bounds of the domain, respectively.
The bin edges satisfy the properties $e_1 < e_2 < \dots < e_{b+1}$, ensuring full coverage of the range, i.e., $\bigcup_{i=1}^{b} [e_{i}, e_{i+1}] = [e_1, e_{b+1}]$.
A continuous value $x$ is mapped to $i$ if $x \in [e_{i}, e_{i+1})$ $\forall i \neq b$ and $\hat{x} = b$ if $x \in [e_{b}, e_{b+1}]$.
The discretized value can also be mapped back onto the continuous range, typically by sampling from the corresponding interval.

\descr{Synthetic Data.}
We focus on differentially private generative models for tabular synthetic data.
More formally, let $A$ be a training algorithm, which takes as input a privacy budget, $\epsilon$, and a training dataset with $n$ records, $D_{train}$.
During the fitting step, $A$ iteratively updates the internal parameters of a generative model to fit the dataset and outputs the trained generative model $G_{\overline{\theta}}$.
In the generation stage, the generative model can be repeatedly sampled to produce a ``fresh'' dataset, denoted as $D_{synth}$.
Typically, synthetic data is evaluated against a separate test dataset, $D_{test}$, drawn from the same distribution, for evaluating downstream tasks.

\descr{Privacy Attacks.}
We use membership inference attacks (MIAs)~\cite{shokri2017membership, hayes2019logan} to estimate the privacy leakage from a trained generative model.
In an MIA, an adversary attempts to determine whether a specific target record was included in the training dataset of the generative model (i.e., $\mathbf{x}_T \in D_{train}$), which closely aligns with the DP definition.
The attack can be formulated as a distinguishing game, where the adversary is presented with either $D_{synth}$, generated by a model trained on $D_{train}$, or $D_{synth}^{\prime}$, generated by a model trained on $D_{train}^{\prime} = D_{train} \setminus \mathbf{x}_T$, and has to predict which training dataset was used.
The adversary's predictions, $Pred = \{pred_1, pred_2, \dots \}$ and $Pred' = \{pred'_1, pred'_2, \dots \}$, are analyzed, and their performance is quantified using the AUC score, which measures their success in distinguishing between the two scenarios.

\section{Evaluation Framework}
We now introduce our experimental evaluation framework and its components, i.e., discretization, generative models, datasets, and metrics.
We also describe how they are integrated across four measurement settings: three focused on utility and one on privacy.

\subsection{Discretization}
\label{subsec:disc}
We experiment with four different discretization strategies, two sampling techniques, and four methods for determining the optimal number of bins.
The DP adaptations (of most) of these methods, which have primarily been studied in non-DP contexts, are introduced later in~\cref{sec:dp_disc}.

\descr{Domain Extraction.}
While it is common to assume that the domain of numerical columns (defined by their minimum and maximum values) is known and/or sourced from public sources,
in the vast majority of %
implementations, these domain values are directly derived from the training data itself~\cite{zhang2017privbayes, ping2017datasynthesizer, vietri2020new, mckenna2021winning, mckenna2022aim, mahiou2022dpart, qian2023synthcity, du2024towards}.

\descr{Discretizers.}
The discretizers work independently on a single continuous column (one-dimension) of the dataset, taking as input the column values, the domain/range of the column, and the desired number of bins ($b$).
We evaluate the following four discretizers:
\begin{itemize}[leftmargin=2em]
	\item[$\bullet$] \textit{\uniform} splits the data domain into $b$ intervals of equal width.
	\item[$\bullet$] \textit{\quantile} distributes data such that each bin contains approximately an equal fraction of data points, specifically $1/b$.
	\item[$\bullet$] \textit{\kmeans} employs a standard k-means clustering algorithm to group the data into clusters and then splits them into non-overlapping intervals.
	\item[$\bullet$] \textit{\privtree}~\cite{zhang2016privtree} is a tree-based method recursively splitting the data domain into subdomains.
  It ensures DP by adding Laplace noise~\cite{dwork2006calibrating} to the count at each step.
  Subdomains are further split if the noisy count exceeds a threshold, $\tau$; otherwise, they become leaves, with bin edges defined by the leaf subdomain.
\end{itemize}

While the number of bins is an input to all discretizers, only the first two guarantee that the data will be partitioned into exactly $b$ distinct bins, while \kmeans and \privtree do not---due to empty clusters or small thresholds, respectively.

\begin{table*}[t]
\small
\centering
\setlength{\tabcolsep}{15pt}
\begin{tabular}{@{}lp{13cm}@{}l@{}}
  \toprule
\hspace*{0.1cm}    Doane's formula~\cite{doane1976aesthetic}   & $b^{\mathbf{DF}} := 1 + \log_2{n} + \log_2{(1 + |\hat{g_1}| / \sigma_{\hat{g_1}})}$,  & \hspace{0.2cm}    \\
                                                & where $\hat{g_1}$ is the estimated skewness of the data and  $\sigma_{\hat{g_1}} = \sqrt{6(n-2)/((n+1)(n+3))}$. The formula aims to improve on previous methods~\cite{sturges1926choice} on non-normal data.                                                                            \\
  \midrule
\hspace*{0.1cm}    Rice Rule, aka                              & $b^{\mathbf{RL}} := 2 \sqrt[3]{n}$.\\
\hspace*{0.1cm}    Scott's Rule~\cite{scott1979on} & \\
  \midrule
\hspace*{0.1cm}    Freedman–Diaconis                           & $b^{\mathbf{FDR}} := (x_{max} - x_{min}) / h$,                          \\
\hspace*{0.1cm}    Rule~\cite{freedman1981on}                  & where the bin with $h = 2 IQR(x) / \sqrt[3]{n}$. The rule optimizes for the width of the bins by using the range and the interquartile range of the data (IQR, the difference between 3rd and 1st quartiles).   \\
  \midrule
\hspace*{0.1cm}    Shimazaki-Shinomoto                         & $b^{\mathbf{SHI}} := (x_{max} - x_{min}) / h$,                                     \\
\hspace*{0.1cm}    choice~\cite{shimazaki2006recipe}           &  where the bin with height that optimizes $\arg \min_{h} (2k - v) / h^2$. \newline Here, $k = 1/b \sum_{i=1}^{b} k_i$, $v = 1/b \sum_{i=1}^{b} (k_i - k)^2$, with $k_i$ representing the number of points in the b-th bin.    \\
  \bottomrule
\end{tabular}
\caption{Strategies to select the optimal number of bins.}
\label{tab:optimal}
\reduce
\end{table*}

\descr{Sampling.}
Discretization strategies must be invertible so that the discrete space of generated synthetic data can be post-processed back into the original numerical domain of the training data.
Typically, this is achieved by sampling continuous values from within the bin ranges corresponding to the generated synthetic records.
We experiment with two sampling strategies:
\begin{enumerate}[leftmargin=2em]
  \item \textit{Uniform} samples uniformly within the defined bin edges.
	\item \textit{Mixture}, inspired by~\cite{patki2016synthetic}, fits a truncated normal distribution to each bin during the pre-processing stage and subsequently samples from these distributions during post-processing.
\end{enumerate}

\descr{Optimal Number of Bins.}
In~\cref{tab:optimal}, we list various methods for selecting the optimal bin count.
These typically consider various data distribution aspects, e.g., number of records, moments, range.

\subsection{DP Generative Models}
We now review the five DP generative models and a baseline model used in our evaluation.
Broadly, these models rely on the {\em select–measure–generate} paradigm~\cite{mckenna2021winning, mckenna2022simple}, as they: 1) select a collection of (low-dimensional) marginals or a workload of queries, 2) measure them privately with a noise-addition mechanism, and 3) generate synthetic data which is consistent with the measurements.
Two of these models (PrivBayes and MST) use graphical models to choose the marginals, while the remaining three (RAP, GEM, and AIM) are workload-aware, enabling statistical/linear queries to be well preserved in the synthetic data.

\chgTag{C2}{We select the five DP marginal models, as they have been published in top-tier venues, are among the most widely used/cited, and have open-source implementations.
Overall, our experiments are representative of the space, and we are not knowingly leaving out any alternatives that might yield significantly different results.}

\descr{PrivBayes}~\cite{zhang2017privbayes}
relies on a Bayesian network to select $k$-degree marginals by optimizing the mutual information between them, using the Exponential mechanism~\cite{dwork2006our}.
Then, propagating through the network, the model relies on the Laplace mechanism~\cite{dwork2006calibrating} to measure noisy counts and translate them to conditional marginals, which could later be sampled to generate synthetic data.

\descr{MST}~\cite{mckenna2021winning}
forms a maximum spanning tree (an undirected graph) of the underlying correlation graph by selecting all one-way marginals and a collection of two-way marginals.
These marginals are noisily measured via the Gaussian mechanism~\cite{mcsherry2007mechanism}.
Finally, to create new data, the measurements are processed through Private-PGM~\cite{mckenna2019graphical}.

\descr{RAP}~\cite{aydore2021differentially}
iteratively identifies high error queries, projects them into a continuous relaxation of the data domain (an extension of the projection mechanism from~\cite{nikolov2013geometry}), and optimizes them through gradient-based methods.
Each round involves private selection (via a tailored Report Noisy Max mechanism~\cite{dwork2014algorithmic} with Gumbel noise)
and measurements via a Gaussian mechanism.
Finally, randomized rounding is used to map the generated data to the original domain.

\descr{GEM}~\cite{liu2021iterative}
follows a similar iterative approach but uses a neural network, modeled after a GAN's generator~\cite{goodfellow2014generative}, to implicitly learn a product distribution representation over the attributes in the data domain.
The learned distribution can directly be used to respond to marginal queries.
To satisfy DP, it employs the Exponential mechanism for sampling and the Gaussian mechanism for measurements.

\descr{AIM}~\cite{mckenna2022aim}
similar to MST, uses the Private-PGM algorithm to generate synthetic data that preserves noisy measurements.
However, AIM dynamically selects marginals of interest based on their overall importance, in contrast to the static selection method used by MST.
The process of selecting and measuring the relevant marginals (with the Gaussian mechanism) is then repeated iteratively until the privacy budget is exhausted or a stopping condition is met.

\descr{dp-synthpop~\cite{mahiou2022dpart}.}
We select dp-synthpop as a baseline model because, while it typically does not provide high utility, it is very fast to run.
It is a DP extension of synthpop~\cite{nowok06synthpop}; it fits a DP logistic regression model~\cite{chaudhuri2011differentially} to each column sequentially using all previously visited columns as features.
During the generation step, the trained classifiers are sampled probabilistically.

\begin{figure*}[t!]
	\centering
	\begin{subfigure}[t]{0.235\linewidth}
		\includegraphics{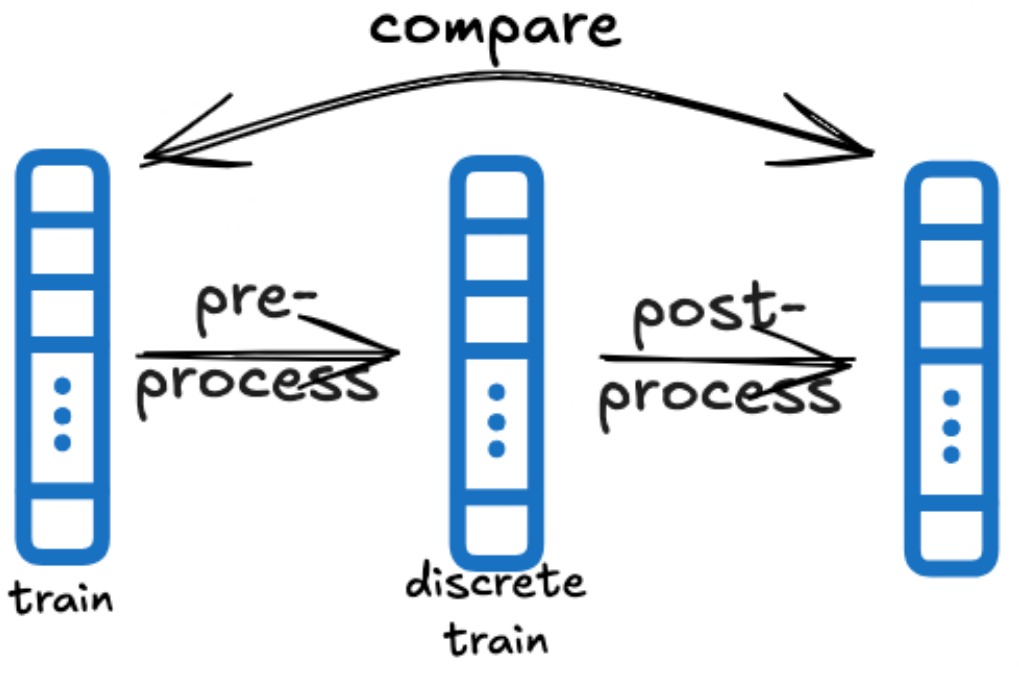}
		 \caption{\footnotesize US1: Utility of DP discretizers preserved on controlled distributions.}
     \label{fig:set1}
	 \end{subfigure}
	 \hfill
	 \begin{subfigure}[t]{0.235\linewidth}
		 \includegraphics{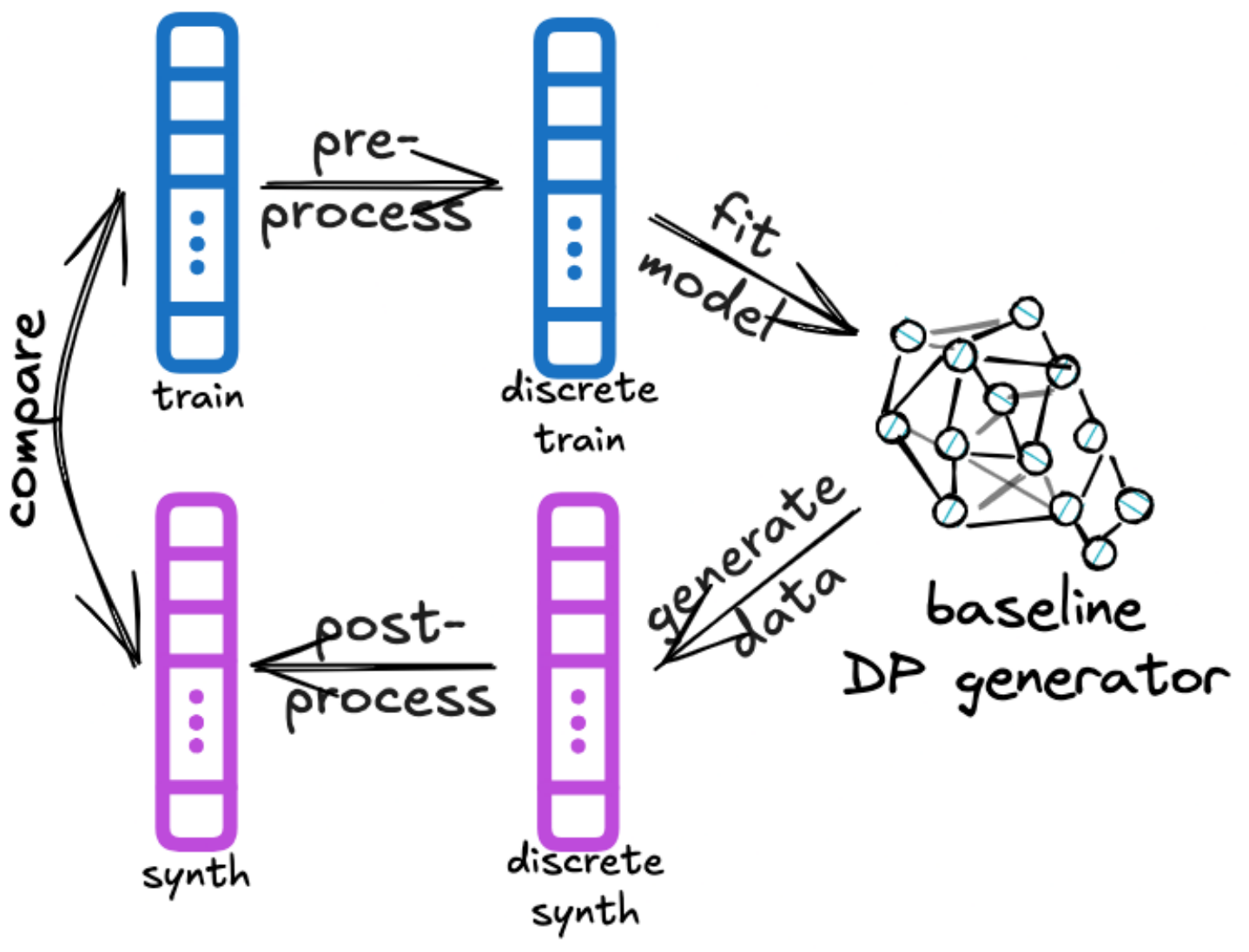}
		 \caption{\footnotesize US2: Utility of DP discretizers and baseline DP generative model preserved on controlled distributions.}
     \label{fig:set2}
	 \end{subfigure}
	 \hfill
   \begin{subfigure}[t]{0.235\linewidth}
     \includegraphics{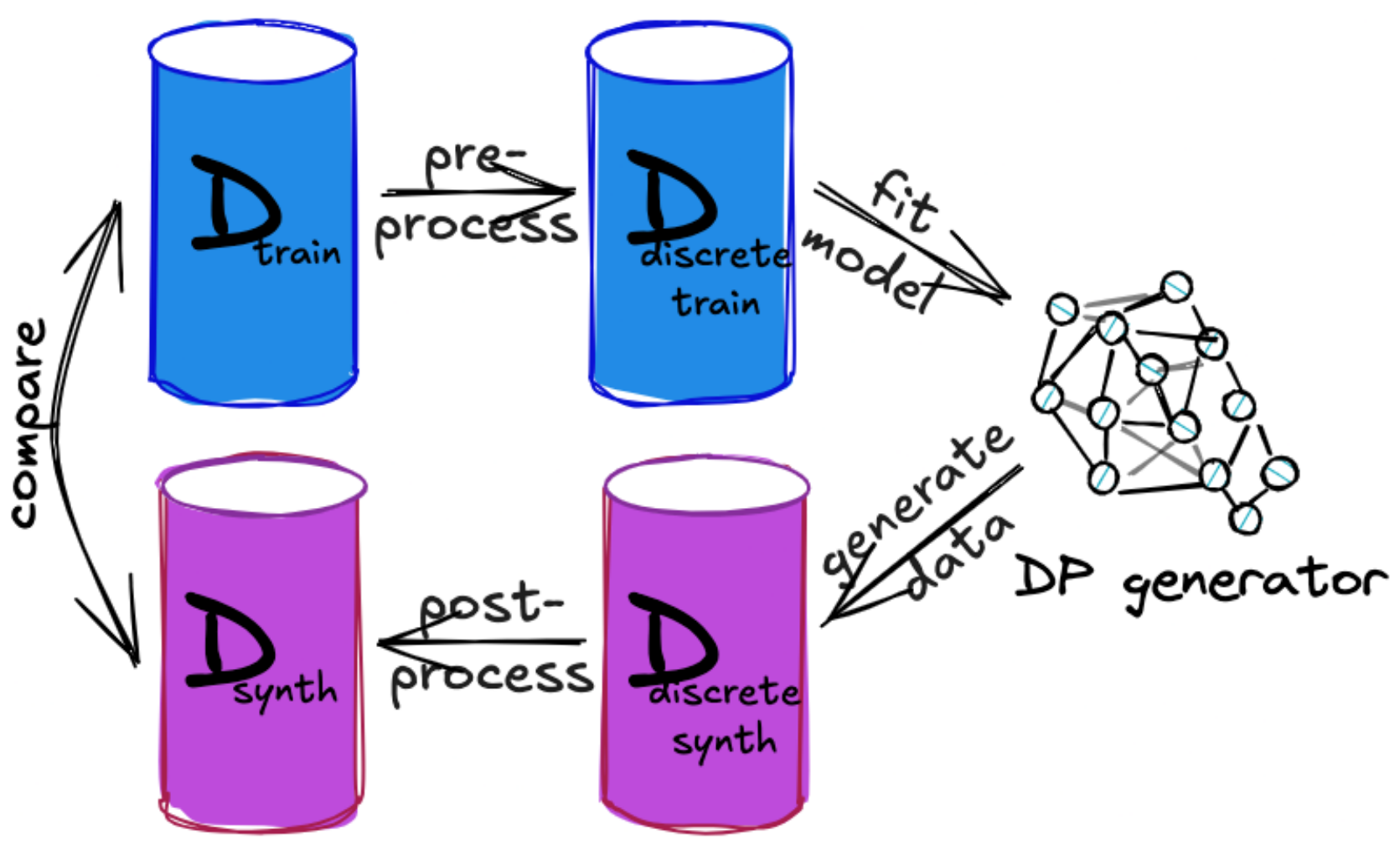}
     \caption{\footnotesize US3: Utility of DP discretizers and DP generative models preserved on real datasets.}
     \label{fig:set3}
   \end{subfigure}
	 \hfill
    \begin{subfigure}[t]{0.235\linewidth}
      \includegraphics{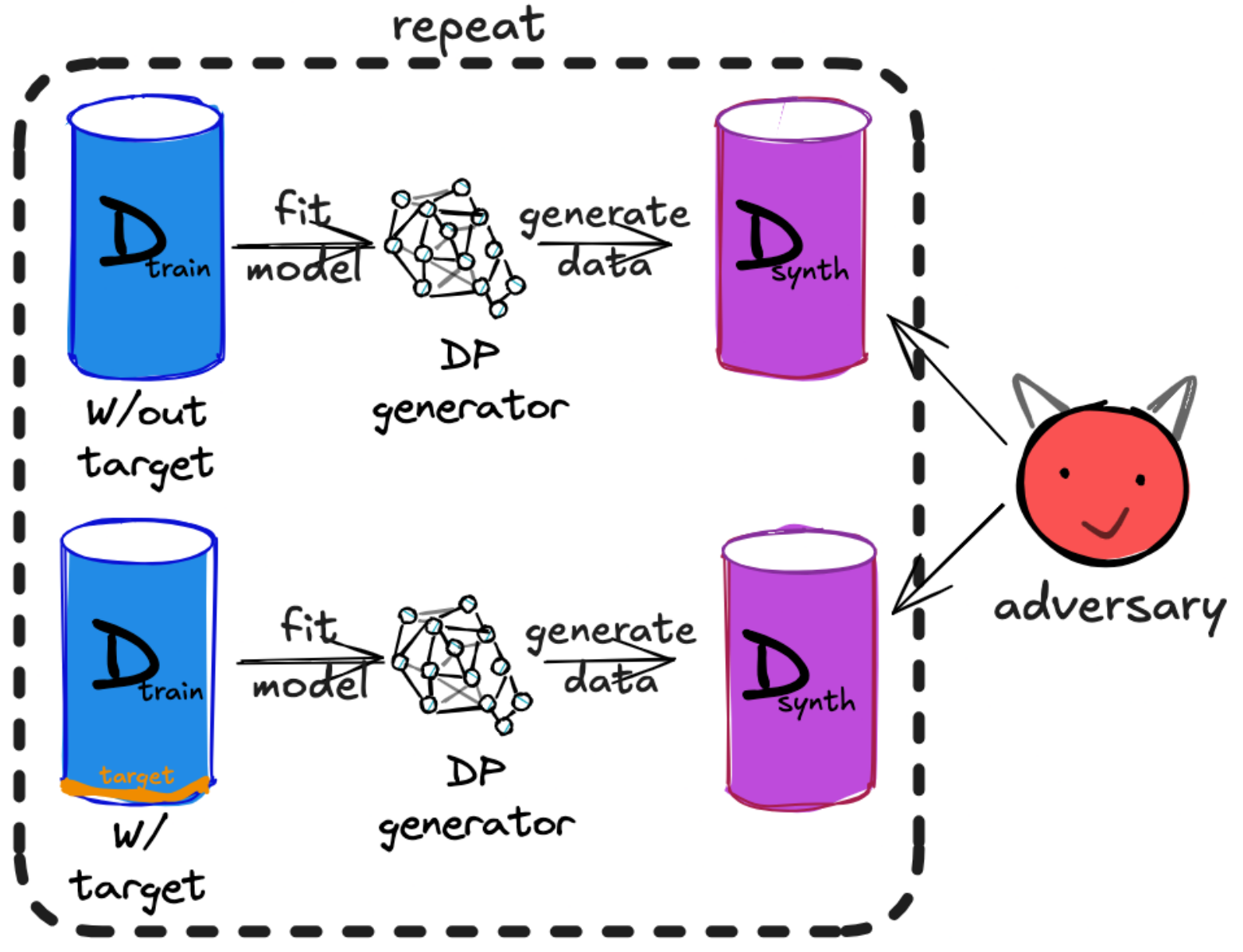}
      \caption{\footnotesize PS1: Data domain's contribution to the privacy of DP generative models.}
      \label{fig:set4}
    \end{subfigure}
	 \caption{The four experimental settings (three focused on utility and one on privacy) used in our measurement study.}
	 \label{fig:set}
   \reduce
\end{figure*}

\subsection{Metrics}
We evaluate the utility of the generated synthetic datasets using standard metrics applied to individual columns, column pairs, and downstream tasks.

\descr{Record Similarity (RS).}
RS is used when comparing two sets of records with a one-to-one correspondence.
It is computed as the average normalized L1 distance: %
\begin{equation}
\scalebox{\sscale}{%
$RS = 1 - \dfrac{\sum_{i=1}^{n} |D_{train}^{i} - D_{synth}^{i}|}{n}$} %
\end{equation}

\descr{Maximum Percentile Distance (MPD).}
MPD measures the maximum normalized L1 distance between the percentiles of two distributions, identifying their ``farthest point.''
Percentiles (0\% to 100\%) are computed after normalizing both columns:
\begin{equation}
\scalebox{\sscale}{%
$MPD = 1 - max_{p=1}^{100}\left|q(D_{train}, p) - q(D_{synth}, p)\right|$}
\end{equation}

\descr{Discriminator Similarity (DS).}
DS trains a classifier to distinguish between real and synthetic records, using a held-out test set to predict probability scores.
The score is calculated as the average absolute distance from 0.5 across all predicted probabilities:

\begin{equation}
\scalebox{\sscale}{%
$DS = \dfrac{1 - 2 * \sum_{i=1}^{n}(|0.5 - p_i|)}{n}$}
\end{equation}

\descr{Query Similarity (QS).}
QS evaluates the preservation of multivariate distributions by applying random multi-dimensional queries (1, 2, and 3d) to both real and synthetic data, and calculating the average Jaccard distance of the resulting counts:

\begin{equation}
\scalebox{\sscale}{%
$QS = \dfrac{1}{|Q|} \sum_{q \in Q} \dfrac{min(q(D_{train}), q(D_{synth}))}{max(q(D_{train}), q(D_{synth}))}$}
\end{equation}

\descr{Correlation Similarity (CS).}
CS measures how well a model preserves pairwise correlations by computing the correlation matrices for both real and synthetic data and then calculating the average Jaccard distance between the normalized matrices:

\begin{equation}
\scalebox{\sscale}{%
$CS = \dfrac{1}{d^2} \sum_{i = 1}^{d}\sum_{j = 1}^{d} \dfrac{min(corr_{train}^{ij}, corr_{synth}^{ij})}{max(corr_{train}^{ij}, corr_{synth}^{ij})}$}
\end{equation}

\descr{Predictive Utility (PU).}
PU measures the performance of classifiers trained on real and synthetic data by comparing their F1 scores on a held-out dataset.
The utility score is calculated as the ratio of the synthetic to real F1 scores, using logistic regression with default hyperparameters:
\begin{equation}
\scalebox{\sscale}{%
$PU = \dfrac{F1_{synth}}{F1_{real}}$}
\end{equation}

\subsection{Datasets}
\label{subsec:data}
We conduct our experiments on five one-dimensional datasets with various controlled distributions along with three real higher-dimensional datasets commonly used for machine learning tasks.

\descr{Controlled Distributions.}
We create five datasets, namely, monotone, normal, beta, skewed, mixture, and imbalanced, reported in~\cref{tab:con}.
(Also see~\cref{subsec:q1} for visual details.)
We also include an additional dataset, uniform, solely for visualization purposes.
The datasets are clipped and rescaled to fit within the $[-10, 10]$ domain, and include a varying number of records, i.e., 500, 1k, 10k, and 100k.

\begin{table}[t!]
	\small
	\centering
	\begin{tabular}{l|l}
	  \toprule
		  \textbf{Controlled Distribution}	& \textbf{Formula}	              \\
		\midrule
		  \textbf{Uniform}		              & $U(0, 1)$                       \\
		  \textbf{Monotone}					        &	$\sqrt{U(0, 1)}$                \\
		  \textbf{Normal}				            &	$N(0, 1)$                       \\
		  \textbf{Beta}			                &	$Beta(2, 14)$                   \\
      \textbf{Mixture}			            & $0.5 N(2, 1) + 0.5 N(9, 1)$     \\
      \textbf{Imbalanced}			          & $0.95 N(0, 1) + 0.05 N(10, 1)$  \\
	  \bottomrule
  \end{tabular}
	\caption{Overview of the controlled distributions.}
	\label{tab:con}
  \reduce
\end{table}

\descr{Real Datasets.}
We use common real datasets, i.e., Adult, Gas, and Wine from UCI's ML Repository~\cite{dua2017data}, all of them having associated binary classification tasks -- see~\cref{tab:real}.

\subsection{Experimental Settings}
\label{subsec:settings}
We introduce four experimental settings, three focused on utility and one on privacy, as visualized in~\cref{fig:set}.
In all utility settings, we fit three generative models, generate three synthetic datasets per trained model, and report aggregated metrics based on the generated 9 synthetic datasets.
\chgTag{C3}{We average across the metrics to provide a concise unbiased score, which covers various use cases and enables straightforward comparison across discretizers/models.}

\descr{Capturing Controlled Distributions (US1).}
We measure how much utility DP discretizers preserve in controlled settings (cf.~\cref{subsec:q1}).
The goal is to isolate their performance under controlled conditions.
To do so, we discretize the input datasets using the four discretizers and sample uniformly from each bin, assuming a known data domain.
The entire privacy budget is spent in the discretization step and utility is measured by averaging all applicable one-dimensional metrics -- record similarity, maximum percentile distance, discriminator similarity, and 1d query similarity.
We use the five controlled distributions and their size variations.

\descr{Modeling Controlled Distributions (US2).}
Next, we evaluate the utility retained when baseline DP generative modeling is applied alongside DP discretization in controlled settings (as explored in~\cref{subsec:q2} and~\ref{subsec:q3}).
This simulates the simplest generative modeling scenario for one-dimensional data.
The datasets are discretized using the four discretizers, and the distributions are modeled using the Laplace mechanism, as done by dp-synthpop and PrivBayes when applied to one-dimensional data.
The privacy budget is split, with 10\% allocated to discretization and 90\% to modeling.
Unless stated otherwise, we assume the data domain is known (i.e., not extracted from the data) and sample uniformly from each bin.
Utility is measured using the same metrics as in US1, except for record similarity.
We use the same five controlled distributions as in US1.

\begin{table}[t!]
	\small
	\centering
	\begin{tabular}{l|rrr}
		\toprule
			\textbf{Dataset}	    & \textbf{\#Records} & \textbf{\#Numerical} & \textbf{\#Categorical}	  \\
      	                    &                    & {\bf Columns}        & {\bf Columns}             \\
		\midrule
			\textbf{Adult}        & 48,842		          & 6                     & 8                       \\
			\textbf{Gas}					& 36,733              & 12                    & 0                       \\
			\textbf{Wine}				  & 4,898               & 11                    & 0                       \\
	 \bottomrule
	\end{tabular}
	\caption{Overview of the datasets used in our evaluation.}
	\label{tab:real}
  \reduce
\end{table}

\descr{Modeling Real Datasets (US3).}
We also assess the utility of DP generative models combined with DP discretization \chgTag{C3}{on real datasets (used in~\cref{subsec:q2},~\ref{subsec:q3}, and~\ref{subsec:q4}) to simulate realistic end-to-end DP generative model deployments and test whether our findings in US1/US2 generalize.}
The datasets are preprocessed using the four discretizers and used with six DP generative models.
As in previous work, we apply the models with their default hyperparameters~\cite{tao2021benchmarking, ganev2024graphical}.
The privacy budget is distributed 10\% to discretization, divided equally across columns, and 90\% to modeling.
Unless otherwise specified, the data domain is assumed to be known (we use DP extraction in~\cref{subsec:q4}), and uniform bin sampling is used (we compare uniform and mixture sampling in~\cref{subsec:q2}).
We measure utility with all listed metrics (without record similarity).
For this setting, we use the three real datasets: Adult, Gas, and Wine.

\descr{Privacy Leakage (PS1).}
In the privacy setting, we evaluate the domain's contribution to overall privacy leakage within the end-to-end DP pipeline using MIAs (cf.~\cref{subsec:q4}).
We assess three domain extraction strategies: assuming a given domain, extracting it directly from the input data, or extracting it with a DP.
Due to computational constraints, we focus on four discretizers combined with two generative models, i.e., PrivBayes and MST, whose privacy properties have been extensively studied~\cite{annamalai2024what, golob2025privacy}.

We use the GroundHog~\cite{stadler2022synthetic} \chgTag{C7}{and Querybased~\cite{houssiau2022tapas}} membership inference attacks, which are arguably the most popular attack targeting synthetic data generation~\cite{annamalai2024what, ganev2025elusive}.
First, we identify a vulnerable record as the target, selecting the data point furthest from all others in the training set~\cite{meeus2023achilles} and ensuring it lies outside their domain.
Next, we train two sets of shadow models (200 models each): one trained on the full dataset, including the target record, and the other excluding it.
The discretization privacy budget is allocated based on the domain extraction strategy: 100\% to discretization when the domain is provided or extracted non-privately, and split 50/50 between domain extraction and discretization when the domain is extracted with DP.
We then generate synthetic datasets.
\chgTag{C7}{For GroundHog,} we extract statistical features (minimum, maximum, mean, median, and standard deviation) from each column of the synthetic data, \chg{while for Querybased, we extract the counts of all possible subsets of columns of the synthetic data that match the corresponding values in the target record.}
Half of these features are used to train a classifier, and the AUC score is evaluated and reported on the remaining data.
We use a single dataset, Wine.

\section{DP Discretization in End-to-End DP Generative Pipeline}
\label{sec:dp_disc}
In this section, we introduce the Differentially Private (DP) discretization components used in place of the non-DP versions introduced in~\cref{subsec:disc}, which ultimately allows for the construction of an end-to-end DP generative pipeline.
Specifically, we present DP variants for domain extraction, three discretizers (\uniform, \quantile, \kmeans), two sampling methods (uniform and mixture), and a strategy for selecting the optimal number of bins.
Throughout, we use primitives of two well-known open-source libraries, namely, Harvard's OpenDP~\cite{opendp2021smartnoise} and IBM's Diffprivlib~\cite{holohan2019diffprivlib}.

\descr{DP Domain Extraction.}
We implement the algorithm by Desfontaines~\cite{desfontaines2020lowering} to estimate the domain of numerical data given a privacy budget $\epsilon$.
It derives bounds using a noisy histogram over an exponential range $[-2^{m}, 2^{m}]$, with $m$ typically set to 32.
The bounds are determined by iteratively reducing a threshold until at least one bin exceeds it, using the highest and lowest bin edges above the threshold serving as the domain bounds.\footnote{Inspired by OpenDP, see \url{https://github.com/opendp/smartnoise-sdk/blob/main/sql/snsql/sql/_mechanisms/approx_bounds.py}.}
Note that this method may fail if $\epsilon$ is too small; %
however, since this does not affect the conclusions drawn from our measurement study, addressing this potential issue is left for future work.

\descr{DP Discretization.}
We use the following methods to have the four discretizers satisfy DP (note that the data domain, privacy budget, and number of bins are provided as input to all discretizers):

\begin{itemize}[leftmargin=2em]
	\item \textit{\chg{(DP)}-\uniform} does not consume any privacy budget and determines bin edges only relying on the provided data domain.
	\item \textit{DP-\quantile} splits $\epsilon$ evenly across a given number of bins, with each quantile calculated using $\epsilon/b$.
  We use the method proposed by \citet{smith2011privacy}, which samples quantile values from a discrete distribution.
  Each $q_i$ is computed as $(x_{i+1} - x_i) * \exp(-\epsilon |i - \alpha n|)$, where $x_i$ is the value at index $i$ in the sorted dataset and $\alpha$ is the target quantile. %
	\item \textit{DP-\kmeans} is based on \citet{su2016differentially}, which adds Geometric noise~\cite{ghosh2009universally} to the counts of the nearest neighbors for cluster centers and Laplace to the sum of values per dimension.\footnote{Inspired by Diffprivlib, see \url{https://github.com/IBM/differential-privacy-library/blob/main/diffprivlib/models/k_means.py}.}
	\item \textit{\privtree}~\cite{zhang2016privtree} is inherently DP.
  The threshold parameter $\tau$ is set to ${1}/{b}$, making $b$ an upper limit for the actual number of bins produced.
\end{itemize}

\descr{DP Sampling.}
We explore two strategies for DP sampling, which are compatible with all discretization strategies.
More precisely:

\begin{itemize}[leftmargin=2em]
	\item \textit{\chg{(DP)}-Uniform} sampling does not use $\epsilon$ as it samples uniformly from the corresponding bin edges.
	\item \textit{DP-Mixture} models each bin as a truncated normal distribution, with mean and standard deviation computed under DP.
\end{itemize}

\descr{DP Selection of the Optimal Number of Bins.}
Finally, we optimize the Rice rule's formula ($b^{\mathbf{RL}}$) in the context of DP, as the alternatives ($b^{\mathbf{DF}}$, $b^{\mathbf{FDR}}$, and $b^{\mathbf{SHI}}$) depend on aspects of the data distribution, requiring additional privacy budget expenditure.
We incorporate a correction term, taking into account the available privacy budget $\epsilon$, and define $b^{\mathbf{RL-OPT}} := \frac{2 \sqrt[3]{n}}{1 + exp(-\epsilon)}$.
The correction term aims to reduce the number of bins for small $\epsilon$.

\section{Experimental Evaluation}
\label{sec:exp}
In this section, we present our experimental evaluation aimed at addressing the four research questions introduced in~\cref{sec:intro}.

\begin{figure}[t!]
	\centering
	\includegraphics[width=\mywidth\linewidth]{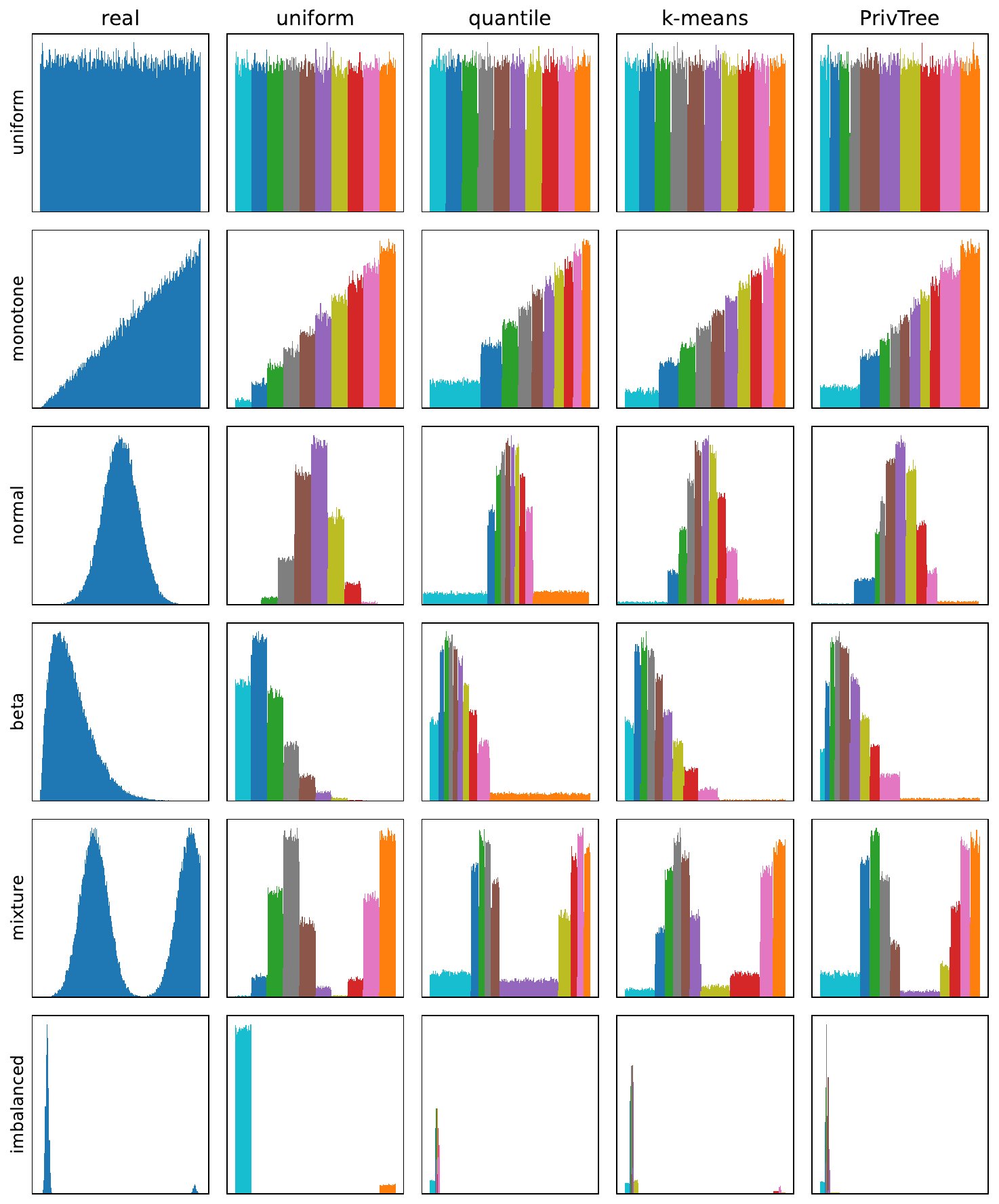}
	\caption{Six controlled distributions and their discretization by four DP discretizers with 10 bins and $\epsilon=\infty$.}
	\label{fig:data}
	\reduce
\end{figure}

\subsection{RQ1: Discretization and Capturing Controlled Distributions}
\label{subsec:q1}
We begin by discussing the discretization of controlled distributions in US1 (cf.~\cref{subsec:settings}) and measure how much utility is preserved.

\descr{Motivating Examples.}
Before diving into the details of our experiments, we first illustrate, visually, the effects of discretization on the six controlled distributions -- see~\cref{fig:data}.
The differences between discretization strategies are clear in how they partition the space, particularly in the placement and spacing of bin edges, as well as the distribution of data across bins.

In~\cref{fig:clustering} and~\ref{fig:classification} in Appendix~\ref{app:exp}, we also present two examples that visually highlight the importance of discretization in simple clustering and classification tasks.
Overall, we observe that discretization can enhance modeling in certain scenarios.

\begin{figure}[t!]
		\centering
		\includegraphics[width=\mywidth\linewidth]{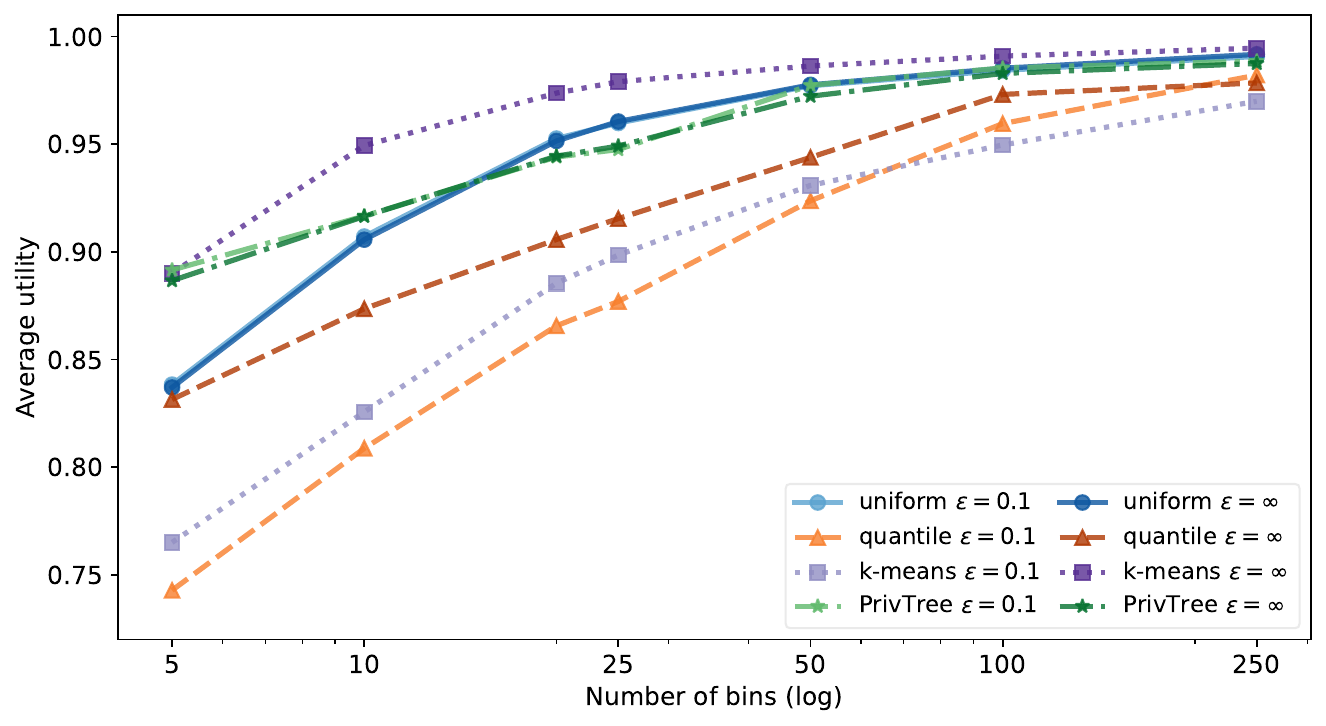}
		\caption{Utility of DP discretizers, averaged across five controlled distributions (US1).}
		\label{fig:rq1_bins}
    \reduce
\end{figure}

\descr{Discretizers Utility (US1).}
Next, we measure how well the four discretizers capture the controlled distributions as per US1 (%
cf.~\cref{fig:set1}).
In~\cref{fig:rq1_bins}, we report the aggregated utility averaged across the datasets.
Utility improves as the number of bins increases for all discretizers.
Remarkably, they approach the maximum achievable score (within 5\%) when 250 bins are used, even under tight privacy constraints ($\epsilon=0.1$).
This suggests that increasing the number of bins effectively offsets the added noise introduced by DP, likely because finer granularity reduces the distance between sampled points and their original values.
At $\epsilon=\infty$, \kmeans is the best-performing discretizer, although it experiences the biggest decline when DP is applied.
By contrast, while \uniform remains unchanged with DP, \privtree also shows minimal change in performance.

\descr{Take-Aways.}
Discretization impacts data representation and, despite being lossy, can benefit downstream tasks under certain conditions.
Utility improves with more bins, effectively countering DP noise when the discretization process is the only step subject to DP.
Uniform and \privtree perform consistently well.

\begin{figure}[t!]
		\centering
		\includegraphics[width=\mywidth\linewidth]{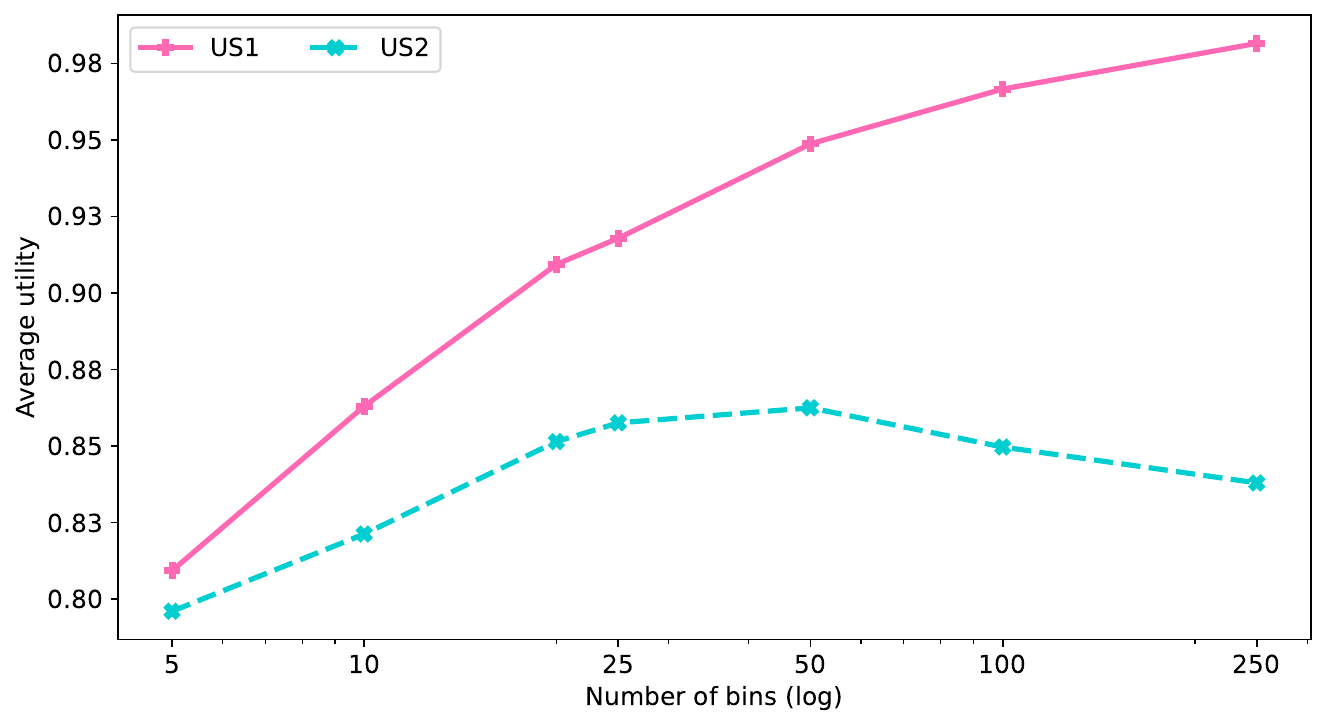}
    \caption{Utility of DP discretizers w/ and w/out baseline DP generative modeling ($\epsilon=0.1$), averaged across five controlled distributions (US1 vs. US2).}
		\label{fig:rq1_rq2_bins}
    \reduce
\end{figure}

\begin{figure}[t!]
  \centering
	\includegraphics[width=\mywidth\linewidth]{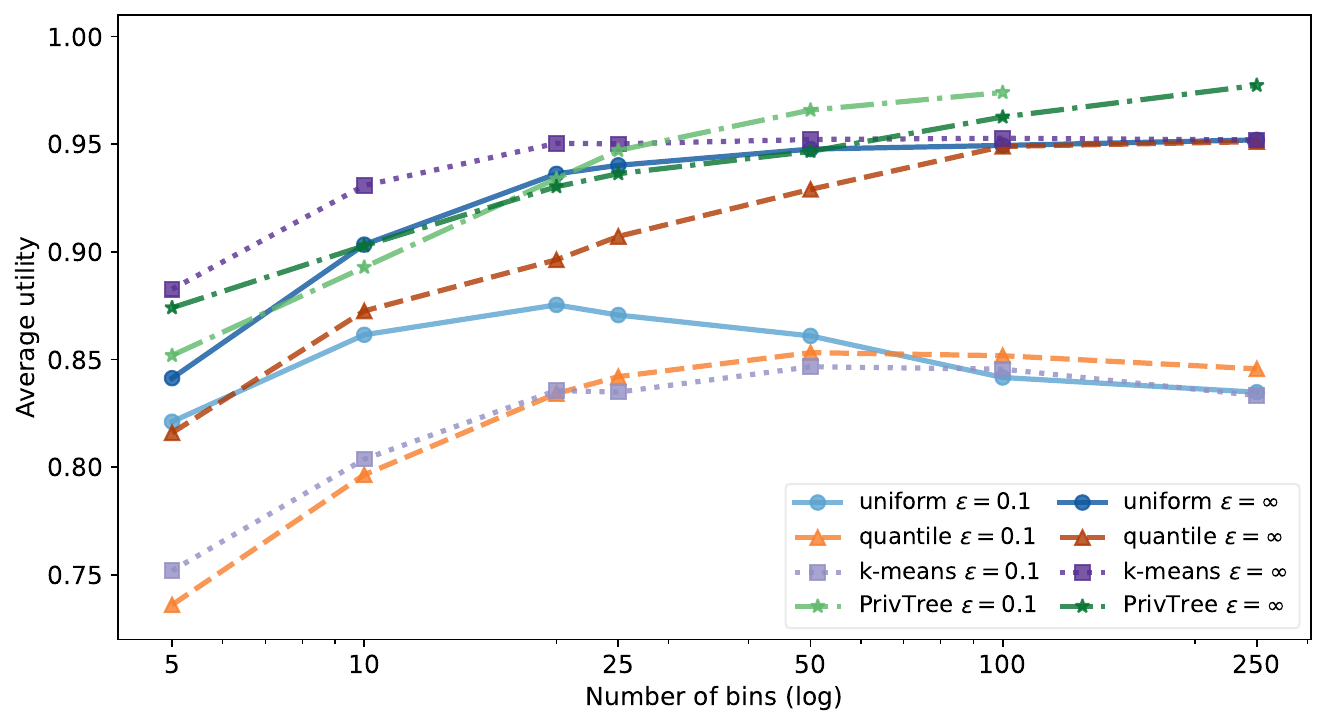}
  \caption{Utility of DP discretizers w/ baseline DP generative modeling, averaged across five controlled distributions (US2).}
	\label{fig:rq2_bins_1d}
  \reduce
\end{figure}

\subsection{RQ2: Discretization and DP Generative Models}
\label{subsec:q2}
Next, we examine the four discretizers in combination with baseline DP generative modeling controlled \chgTag{C3}{one-dimensional} distributions (US2; see~\cref{fig:set2}) to understand their behavior in controlled settings.
We then assess their impact when used with DP generative models on real \chg{high-dimensional} datasets (US3, see~\cref{fig:set3}).

\begin{figure*}[t!]
	\centering
	\includegraphics[width=\mywidth\linewidth]{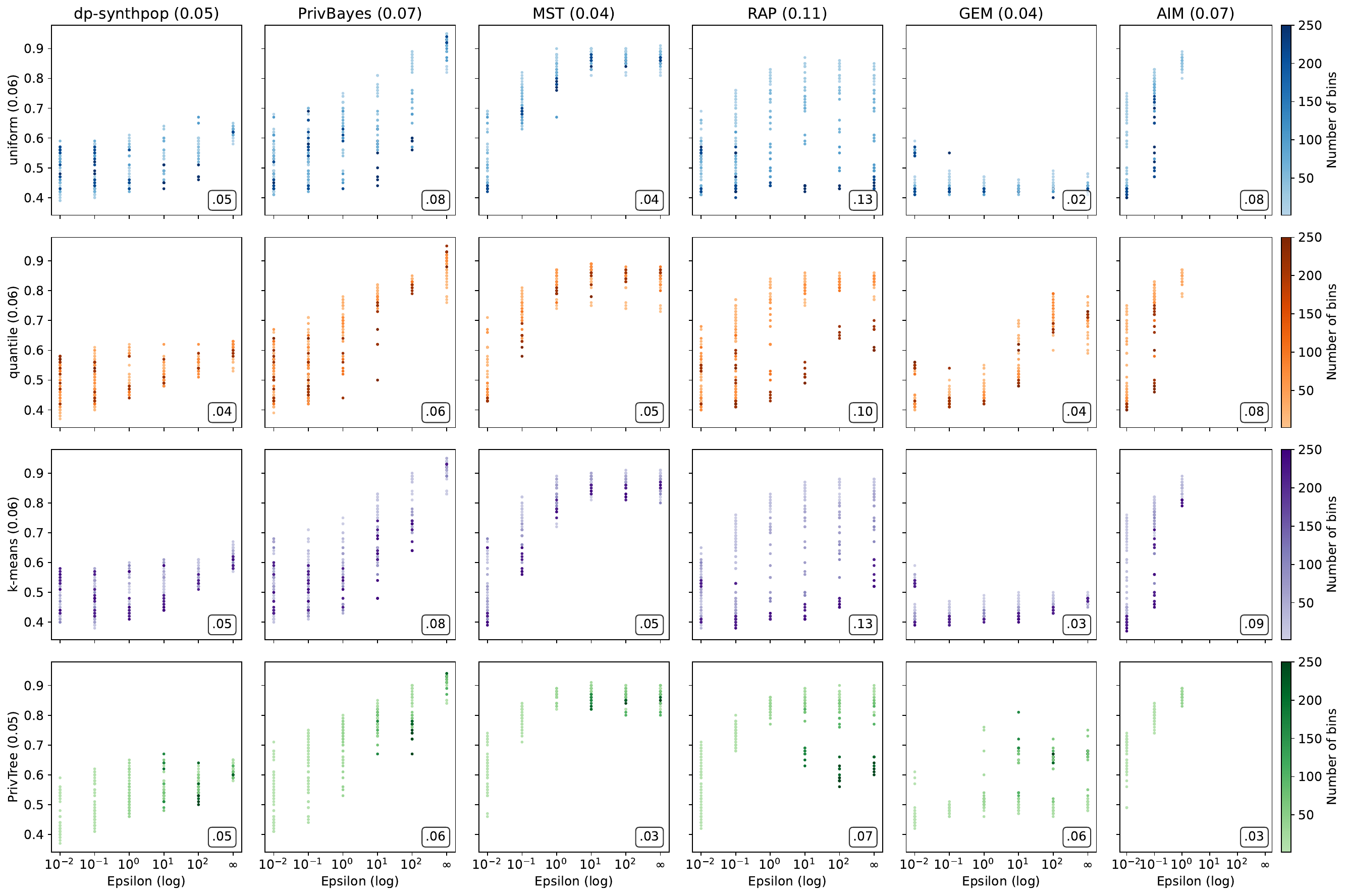}
	\caption{\chgTag{C4}{Utility broken down per DP generative model and DP discretizer over a range of bins (darker dots represent higher bin counts) and $\epsilon$s, while the numbers in boxes/parentheses indicate the utility standard deviation, Gas dataset (US3).}}
	\label{fig:rq2_eps_gas}
  \reduce
\end{figure*}

\descr{Discretizers Utility (US2 vs. US1).}
As per US2, we adopt a simple DP generative model pipeline and compare its performance with US1, which only discretized the data and then sampled back.
We plot an aggregated comparison between US2 and US1 for $\epsilon=0.1$ in~\cref{fig:rq1_rq2_bins}, and a more detailed view of US2 in~\cref{fig:rq2_bins_1d}, showing the four discretizers' behavior across a range of bins and two privacy budgets.
Unlike US1, in US2, increasing the number of bins only helps to a certain point; using more bins beyond it actually harms utility (see~\cref{fig:rq1_rq2_bins}).
This occurs because, for a fixed privacy budget, the noise allocation must be spread across all bins when modeling the distribution.
As the number of bins increases over a certain point, the noise added per bin grows, ultimately degrading the captured distribution and reducing utility.
\chgTag{C5}{This trend generalizes to other $\epsilon$ values as well, with higher budgets improving overall utility and shifting the inversion point to a larger number of bins (for ease of presentation, we only visualize the trend for $\epsilon=0.1$).}

By comparing the discretizers (see~\cref{fig:rq2_bins_1d}), we note that, with $\epsilon=\infty$, which closely mirrors US1, all discretizers continue to improve as the number of bins increases.
However, under DP constraints, most discretizers either plateau or show a slight decline in utility, except for \privtree, which incorporates an internal mechanism to limit the number of bins as $\epsilon$ decreases.
Overall, \privtree demonstrates the most promising performance (except for \kmeans, which outperforms it with up to 25 bins for $\epsilon=\infty$).

\chgTag{C10}{Finally, in~\cref{fig:rq2_bins_1d_dis} in Appendix~\ref{app:us2}, we present the same information as in~\cref{fig:rq2_bins_1d} but disaggregate the three utility metrics in US2.
Although the metric ranges vary, the discretizers show consistent behavior across all metrics, ensuring that each metric is well represented in the final aggregation.}

\descr{Generative Models Utility (US3).}
\cref{fig:rq2_eps_gas} reports the utility results on the Gas dataset, broken down by the four discretizers and six generative models, as per US3 (numbers in \chgTag{C4}{boxes/parentheses} indicate standard deviation).
All generative models are trained with a privacy budget $\epsilon$ in the range \{0.01, 0.1, 1, 10, 100, $\infty$\}, except for AIM, which is limited to $\epsilon \leq 1$ due to computational constraints (it takes over 60 mins for $\epsilon \geq 10$).
We observe that both the choice of discretizer and the number of bins have a significant impact on the performance of all generative models.

Examining their maximum utility per epsilon, we find that only PrivBayes, MST, RAP, and AIM consistently improve with increasing epsilon budgets.
This is a desirable and, arguably, essential quality for any effective DP model.
By contrast, dp-synthpop remains relatively flat, while GEM exhibits flat trends or declines, likely due to its reliance on neural networks, which may require careful hyperparameter tuning.

We also analyze the models' utility variability across different generative models: among the models that do improve (PrivBayes, MST, RAP, and AIM), MST exhibits the lowest variability (0.04), while RAP has the highest variability (0.11).
In most cases, as the DP budget increases, the number of bins becomes less critical to a model's utility, resulting in lower variability.
This suggests that optimizing the number of bins is more crucial for lower epsilon values, where sensitivity to bin count is higher.
RAP is an exception, likely because it directly optimizes over the average error of all queries; increasing the number of bins introduces more queries, complicating the learning process.
By contrast, MST employs an internal compression mechanism that excludes bins with noisy counts below a certain threshold, while AIM adaptively selects the number of queries based on the budget.

Finally, for models with relatively low variability (i.e., PrivBayes, MST, and AIM), we observe a clear trend related to the number of bins.
\chgTag{C5}{At lower epsilon values, the best results are achieved with fewer bins (darker dots are positioned lower).
As epsilon increases, higher numbers of bins yield better results (darker dots shift upward).}
We further explore the importance of finding the optimal number of bins in RQ3 in~\cref{subsec:q3}.

\descr{Optimal Discretizer (US3).}
As already reported in~\cref{fig:delta} (in \cref{sec:intro}), optimizing the choice of discretizer (and the number of bins) can increase the utility of all DP generative models by 9.28\% to 43.54\% compared to the baseline choice, \uniform with 20 bins.
In all cases where the best results are achieved, \privtree emerges as the optimal choice.
Moreover, \privtree consistently outperforms the other discretizers across various factors, namely, privacy budgets, number of bins, generative models, and datasets, achieving the highest performance in 63\% of the cases, while \kmeans ranks second with 16\%.
This confirms previous claims that \privtree is better than \uniform~\cite{tao2021benchmarking}.

\begin{figure}[t!]
	\centering
	\includegraphics[width=\mywidth\linewidth]{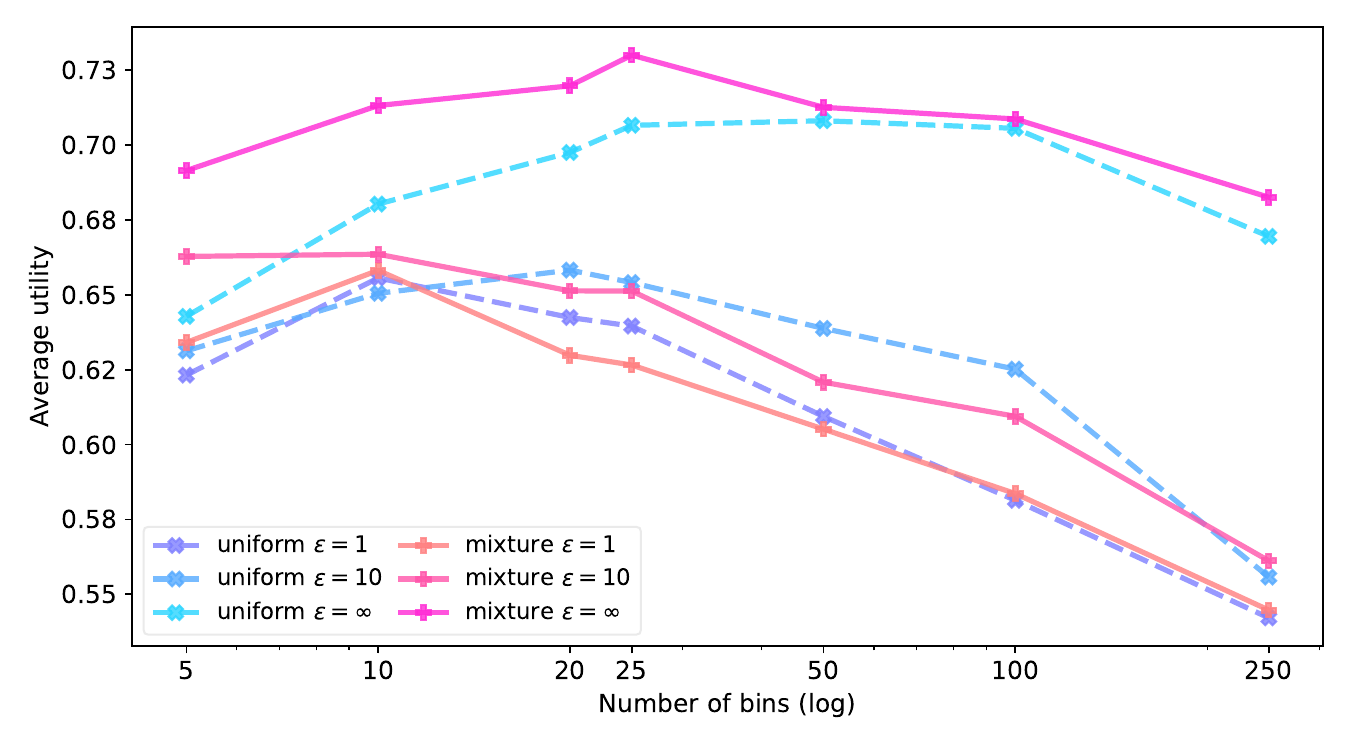}
	\caption{Utility of sampling strategies, averaged across four DP discretizers, six DP models, three datasets (US3).}
	\label{fig:rq2_sampling}
  \reduce
\end{figure}

\privtree's superiority is also visible from~\cref{fig:disc_comp_exp3}, %
where the four discretizers are compared for $\epsilon=1$ \chgTag{C5}{(selected to ease presentation, though the same pattern holds for other $\epsilon$ values)}.
Notably, the overall trend, i.e., increasing the number of bins improves utility up to a point, is consistent with the findings in US2 (cf.~\cref{fig:rq2_bins_1d}).
As for variability, \privtree also achieves the best overall score, albeit by a small margin -- more precisely, 0.05 compared to 0.06 for the other discretizers (cf.~\cref{fig:rq2_eps_gas}).

\descr{Discretizers Sampling (US3).}
In~\cref{fig:rq2_sampling}, we compare the two sampling strategies -- uniform and mixture.
Both sampling methods can be applied to all discretizers in the post-processing step, when converting bin indices back to the original numerical domain.
When using mixture sampling, we allocate 50\% of the discretizer's budget to fitting the truncated normal distributions of the bins.

Using mixture sampling tends to outperform uniform under certain conditions -- high epsilons and low number of bins.
This is perhaps expected, as modeling the distribution of the values within a given bin adds to the expressiveness of the bin representation but at the cost of privacy.
However, once we have enough bins (more than 10), reducing the discretization budget results in uniform sampling outperforming mixture.
This is likely due to inaccurate modeling of mean and standard deviation per bin since the available budget needs to be distributed to too many bins.
Since we tend to achieve the best results with the number of bins more than 10 (see~\cref{fig:disc_comp_exp3} and~\cref{fig:rq2_bins_1d}), we use the uniform sampling by default.

\begin{figure}[t!]
  \centering
  \includegraphics[width=\mywidth\linewidth]{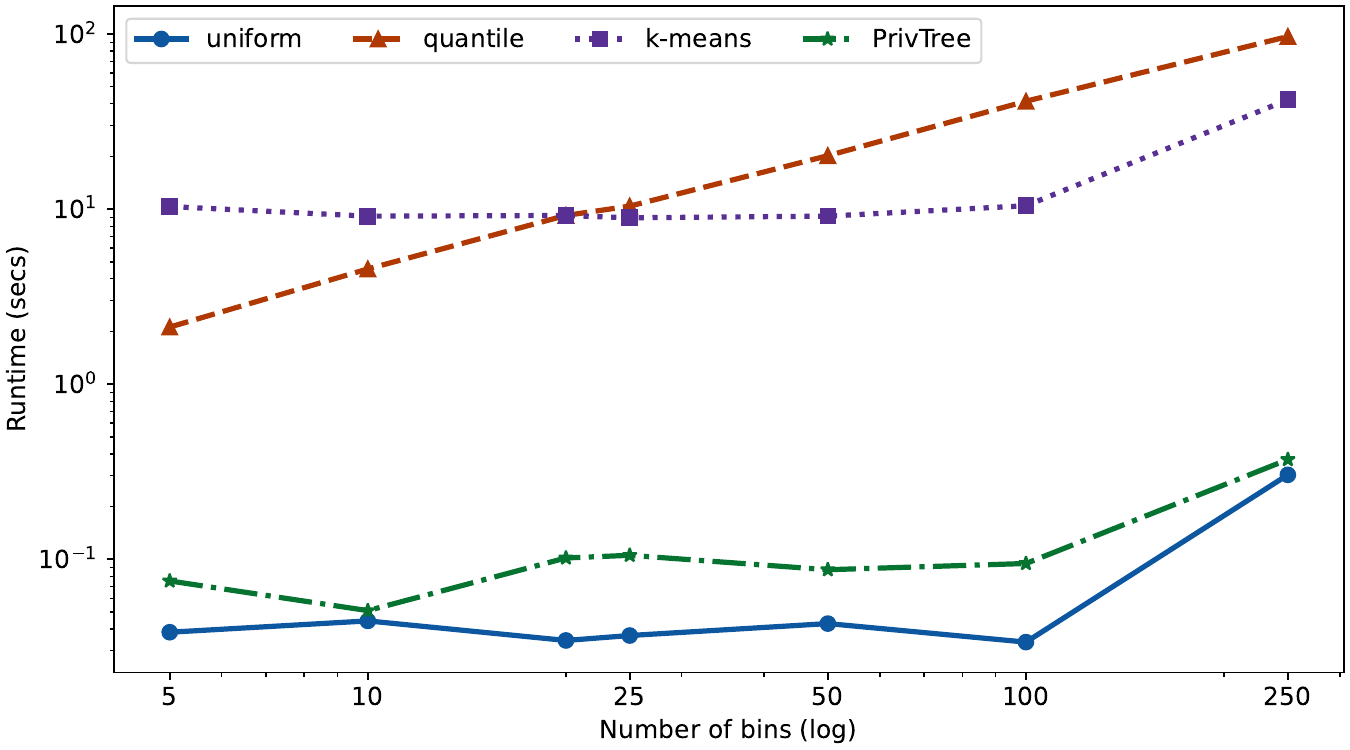}
  \caption{Runtime of DP discretizers, averaged across six DP generative models and three datasets (US3).}
  \label{fig:rq2_runtime}
  \reduce
\end{figure}

\begin{figure}[t!]
	\centering
	\includegraphics[width=\mywidth\linewidth]{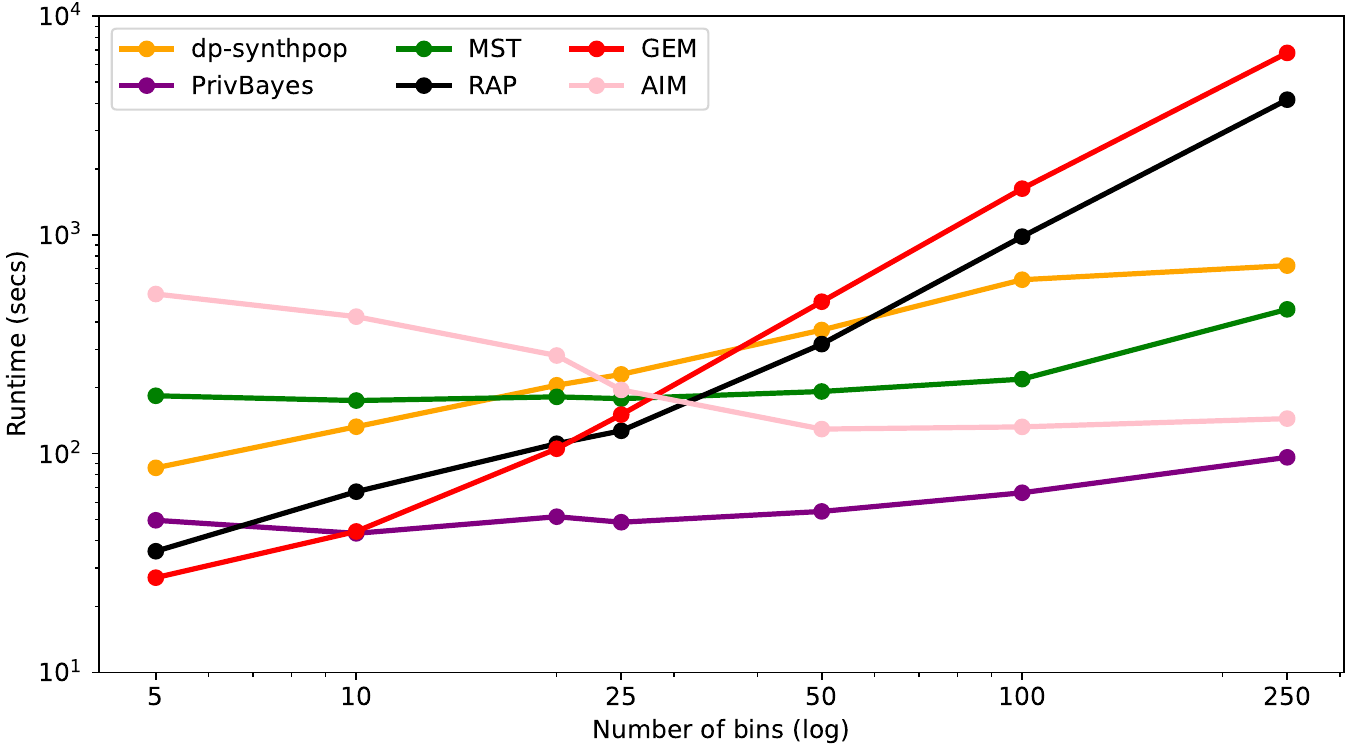}
  \caption{Runtime of DP generative models, averaged across four DP discretizers and three datasets (US3).}
	\label{fig:rq2_runtime_model}
  \reduce
\end{figure}

\descr{Computational Performance (US3).}
We also compare the computational overhead of discretizers and generative models with varying number of bins, reporting runtimes in~\cref{fig:rq2_runtime} and~\cref{fig:rq2_runtime_model}, respectively.
In terms of discretizers' performance, \quantile and \kmeans take longer to fit as the bin count increases, e.g., from 2-10 secs for 5 bins to 40-100 secs for 250 bins.
By contrast, \uniform and \privtree are much faster (under a sec) and, importantly, remain relatively unaffected by bin count, making them highly scalable.

Examining the performance of generative models, dp-synthpop, PrivBayes, and MST slow down linearly as the number of bins increases, yet maintain an acceptable runtime of around 1-7 mins even for 250 bins.
Interestingly, AIM becomes faster with more bins, possibly because it exhausts its budget more quickly when additional bins are available (however, since AIM was trained with $\epsilon \leq 1$, direct comparisons with the other models should be avoided).
Both RAP and GEM slow down exponentially with increasing bins, taking close to 60 mins for 250 bins.

\descr{Take-Aways.}
Our experiments show that DP generative models are highly sensitive to the choice of discretization strategy and bin count, and optimizing them can increase utility by \textasciitilde30\% on average.
While increasing the number of bins initially improves utility, doing so beyond a certain point starts to negatively impact it.
We also find that \privtree consistently outperforms other discretizers in terms of utility, variability, and computational efficiency.

\begin{figure}[t!]
  	\centering
  	\includegraphics[width=\mywidth\linewidth]{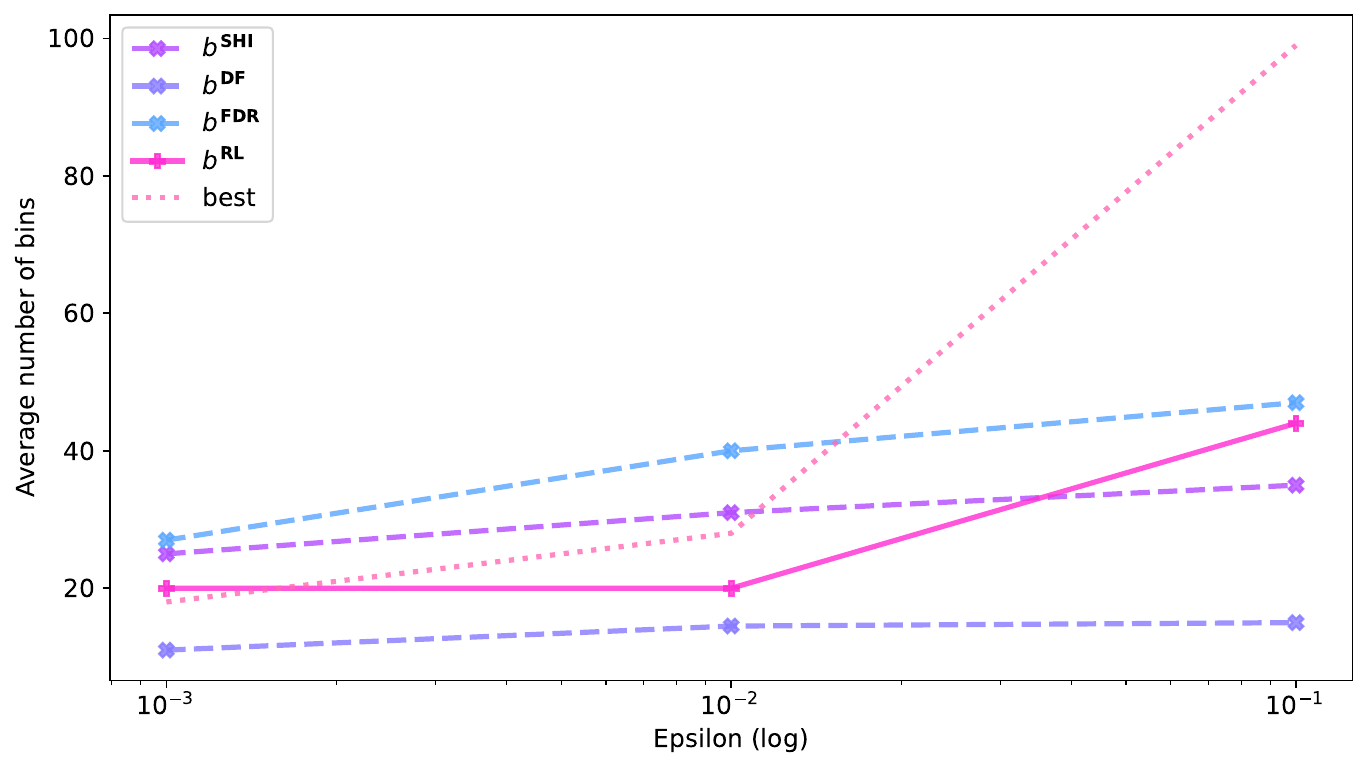}
  	\caption{Utility of strategies for choosing the optimal number of bins w/ baseline DP generative model, averaged across four DP discretizers and five controlled distributions (US2).}
  	\label{fig:rq3_bins_1d}
    \reduce
\end{figure}

\begin{figure}[t!]
		\includegraphics[width=\linewidth]{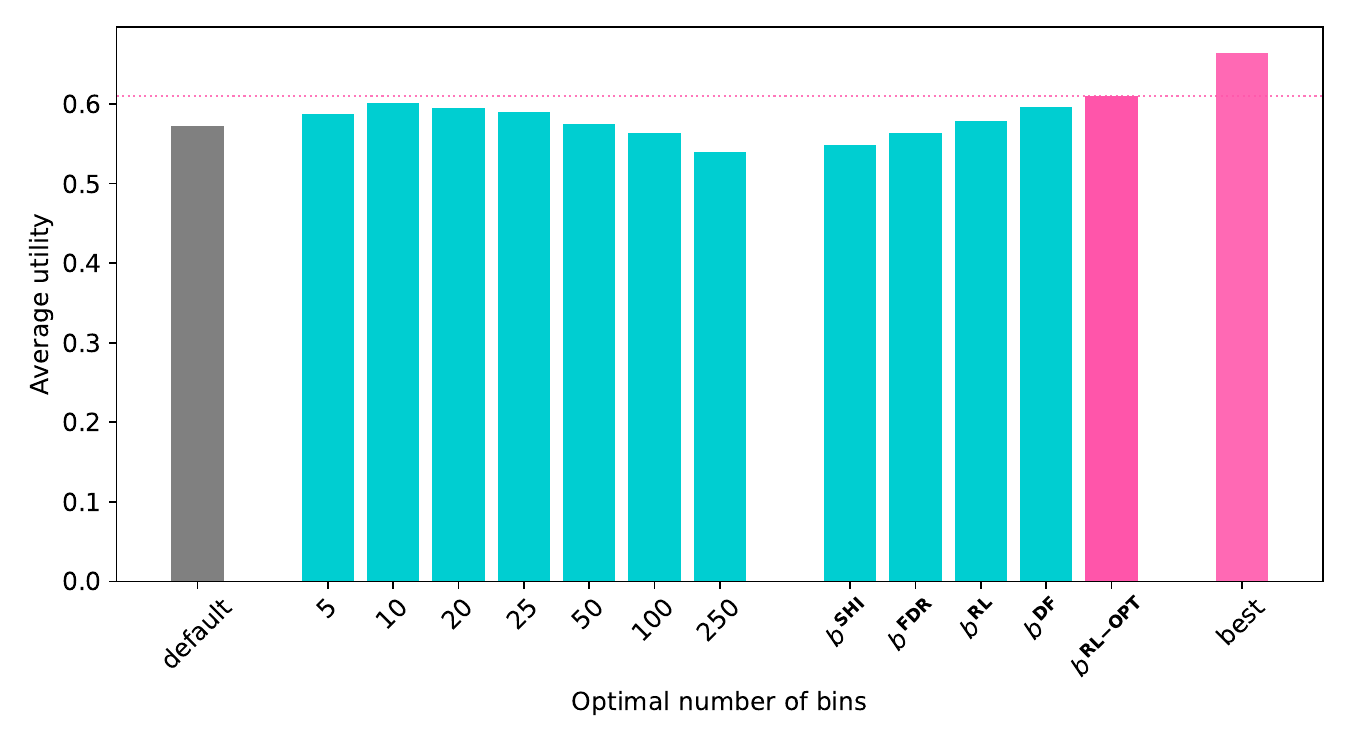}
		\caption{Utility of strategies for choosing the optimal number of bins w/ DP generative models, averaged across four DP discretizers and three datasets (US3).}
		\label{fig:rq3_nbins}
    \reduce
\end{figure}

\subsection{RQ3: Discretization and Optimal Number of Bins}
\label{subsec:q3}
We now set out to explore how to find the optimal number of bins using controlled distributions (US2) and validate our findings with experiments on real datasets (US3).

\descr{Optimized Strategy for Number of Bins (US2).}
In \cref{fig:rq3_bins_1d}, we report the average utility of four strategies for automatically selecting the number of bins -- $b^{\mathbf{DF}}$, $b^{\mathbf{RL}}$, $b^{\mathbf{FDR}}$, and $b^{\mathbf{SHI}}$ (introduced in~\cref{subsec:disc}) -- alongside the number of bins that led to achieving the best utility for the same setting on controlled distributions (US2).
The privacy budget is a key factor in determining the optimal number of bins; as the budget increases, using more bins leads to higher utility -- a trend already noted in~\cref{subsec:q2}.
Among the four bin selection strategies, $b^{\mathbf{SHI}}$ performs best on average.
However, implementing this strategy in a DP manner (i.e., `DPfying' the formula) poses practical challenges, as it relies on numerous unperturbed queries to the data.
\chgTag{C9}{Making these queries DP would likely result in either inaccurate modeling or `wasted' privacy budget.
Similarly, $b^{\mathbf{DF}}$ and $b^{\mathbf{FDR}}$ encounter the same limitations.
By contrast, $b^{\mathbf{RL}}$ provides a better balance between simplicity and performance.}

Therefore, we select $b^{\mathbf{RL}}$ and attempt to optimize it for DP settings because it is data-independent, making it DP without any additional detriment to privacy.
Our optimized version of $b^{\mathbf{RL}}$, denoted as $b^{\mathbf{RL-OPT}}$, takes into account the privacy budget to reduce the number of bins and is introduced in~\cref{sec:dp_disc} \chgTag{C9}{(we also explored more complex correction terms but none yielded better results)}.

\descr{Optimal Number of Bins (US3).}
In~\cref{fig:rq3_nbins}, we report the average utility for all tested bin counts and selection strategies.
\chgTag{C8}{While $b^{\mathbf{RL-OPT}}$ yields the highest utility among all strategies, achieving a 41\% normalized improvement between the default and the best achievable score via exhaustive search, its absolute improvement remains modest.}
This leaves room for future work to develop improved strategies, possibly tailored to specific discretizers.

\chgTag{C9}{Nevertheless, using $b^{\mathbf{RL-OPT}}$ bins for discretization and training a generative model saves privacy budget and reduces computational resources by eliminating the need for extensive hyperparameter search, which is required to achieve the best score.}
By contrast, hyperparameter searches generally require additional privacy budget, making them costly in DP settings~\cite{papernot2022hyperparameter, koskela2024practical}.

\descr{Take-Aways.}
Our optimized bin selection strategy, $b^{\mathbf{RL-OPT}}$, outperforms all tested bin counts and alternative strategies (and we believe there is actually potential for further improvement).

\subsection{RQ4: Discretization and DP Domain Extraction}
\label{subsec:q4}
Finally, we evaluate the impact of DP domain extraction on utility (US3) and privacy (PS1; see~\cref{fig:set4}) across various discretizers and generative models applied to real datasets.
We compare three scenarios:
1) the domain is assumed to be provided and set to that of the full dataset,
2) the domain is extracted directly from the input dataset (without DP), as done in numerous publicly available implementations and libraries~\cite{zhang2017privbayes, ping2017datasynthesizer, vietri2020new, mckenna2021winning, mckenna2022aim, mahiou2022dpart, qian2023synthcity, du2024towards}, and
3) the domain is extracted from the input dataset with DP.
In the first two cases, the entire discretization privacy budget is allocated to the discretizer, while in the third, the budget is split evenly, with 50\% used for domain extraction and 50\% for discretization.

\begin{figure}[t!]
  \centering
  \includegraphics[width=0.8\linewidth]{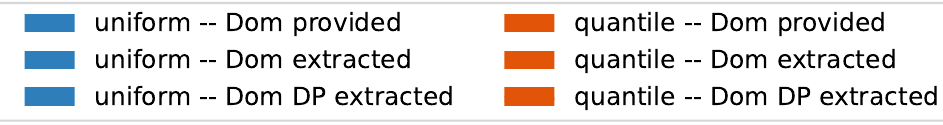}
  \includegraphics[width=0.8\linewidth]{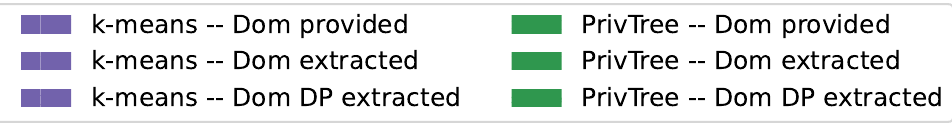}
	\includegraphics[width=0.6\linewidth]{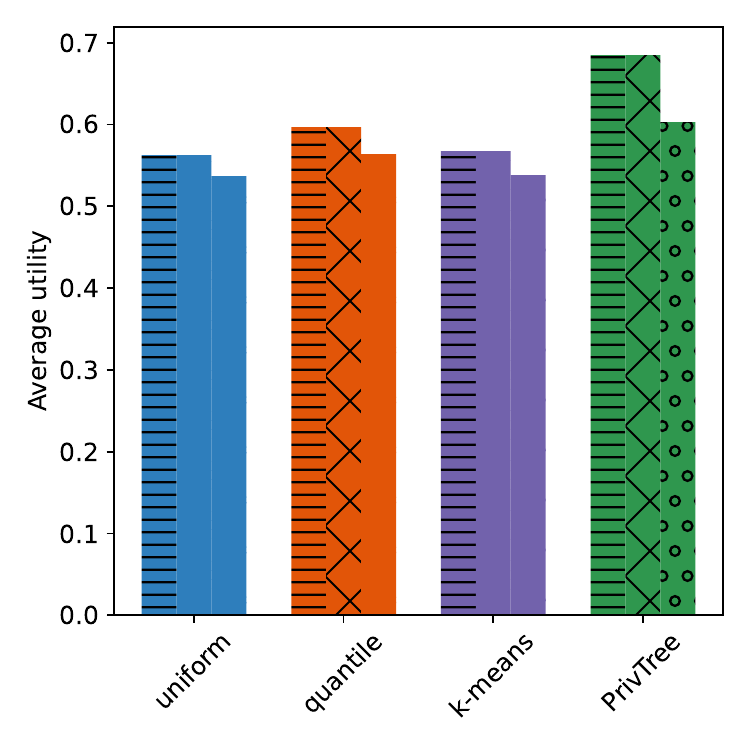}
	\caption{Utility of DP discretizers with \emph{provided} and \emph{extracted} domain (w/ and w/o DP), averaged across six DP generative models and three datasets (US3).}
	\label{fig:rq4_domain}
  \reduce
\end{figure}

\descr{Utility (US3).}
In~\cref{fig:rq4_domain}, we present the average utility across the three domain extraction scenarios for each of the four discretizers, averaged across all models and datasets.
The first two scenarios yield identical results (bars with lines/crosses), as the input dataset matches the full dataset.
Introducing DP domain extraction (bars with circles) reduces utility across all discretizers -- approximately 2.5\% for \uniform, \quantile, and \kmeans, and 8\% for \privtree.
Despite this drop, \privtree still outperforms the others, even when they extract the domain directly without DP.
This experiment underscores the sensitivity of all discretizers to domain specification and highlights the need for future research in this area.

\begin{figure*}[t!]
\includegraphics[width=0.9\linewidth]{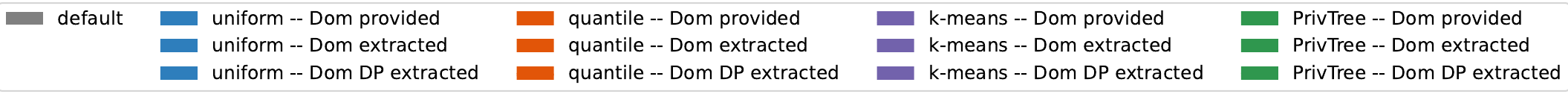}
  \centering
  \begin{subfigure}[t]{\mywidth\columnwidth}
    \includegraphics[width=\mywidth\linewidth]{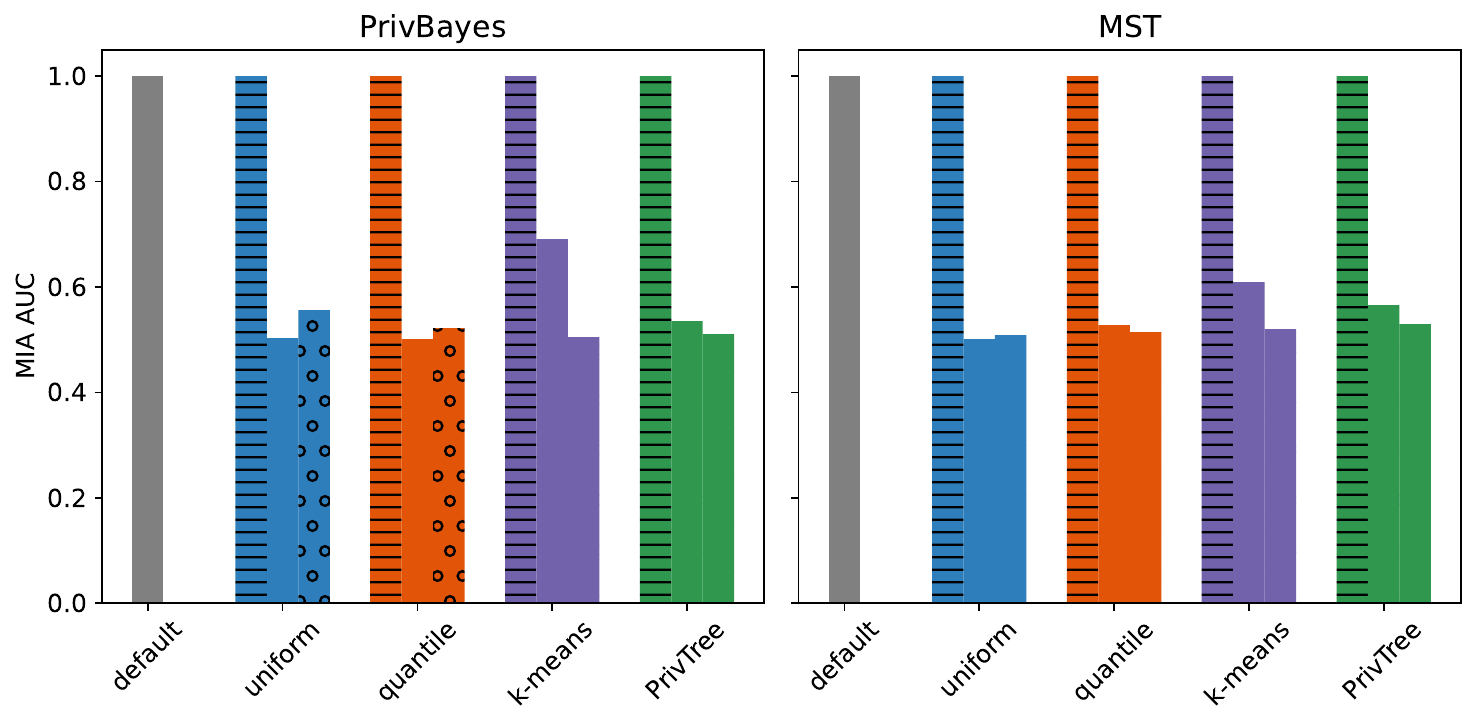}
    \caption{D ($\epsilon=1$), G ($\epsilon=1$), target \emph{outside} domain}
  	\label{fig:rq4_attack}
	\end{subfigure}
  \begin{subfigure}[t]{\mywidth\columnwidth}
    \includegraphics[width=\mywidth\linewidth]{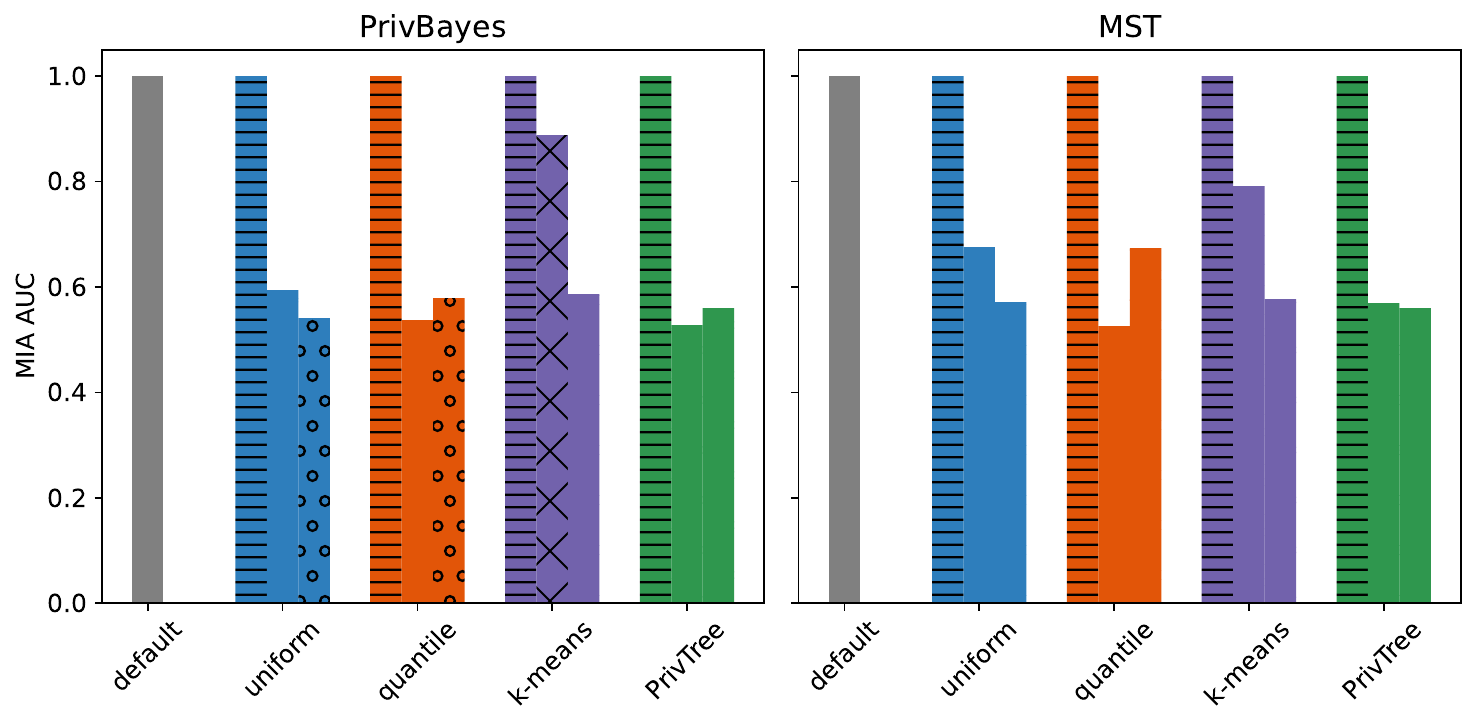}
    \caption{D ($\epsilon=100$), G ($\epsilon=100$), target \emph{outside} domain}
   	\label{fig:rq4_attack_100}
 	\end{subfigure}
  \begin{subfigure}[t]{\mywidth\columnwidth}
    \includegraphics[width=\mywidth\linewidth]{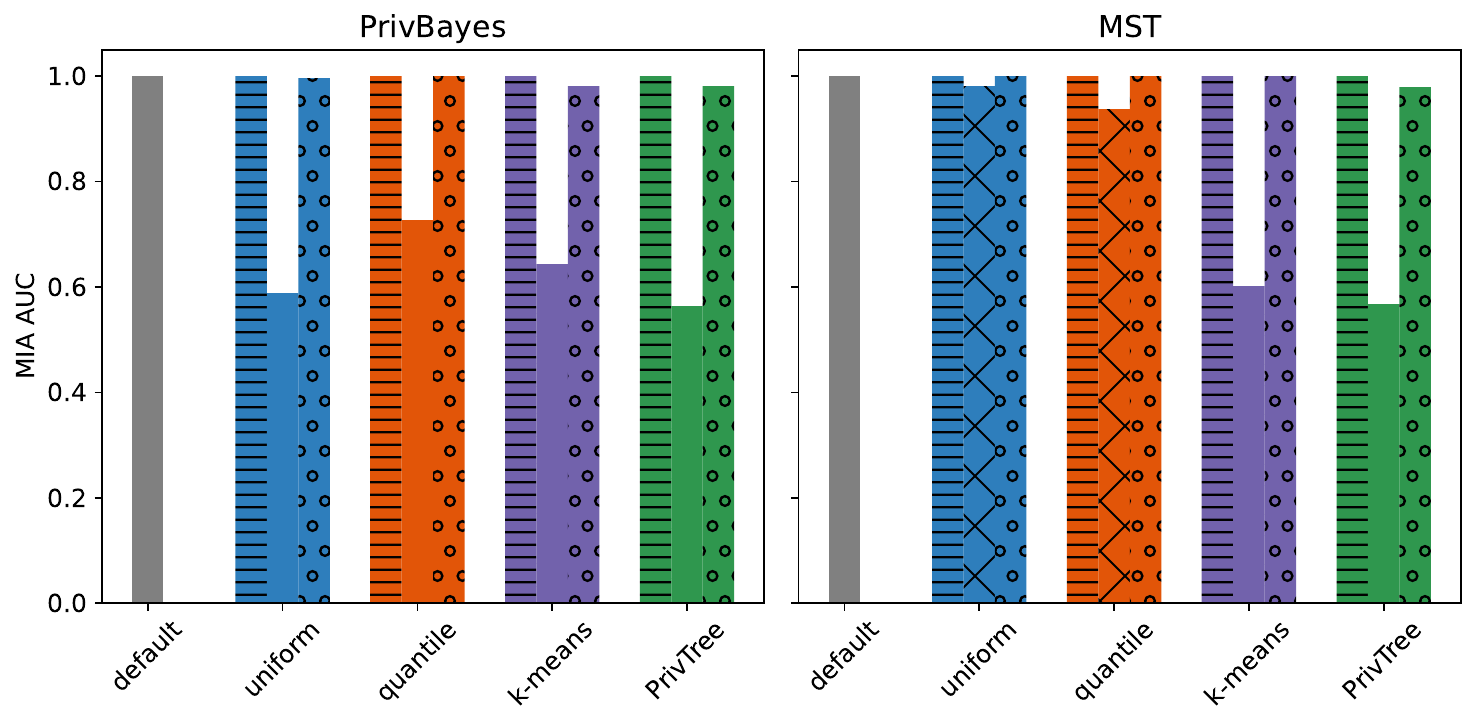}
    \caption{D ($\epsilon=1,000$), G ($\epsilon=1,000$), target \emph{outside} domain}
   	\label{fig:rq4_attack_1000}
  \end{subfigure}
  \begin{subfigure}[t]{\mywidth\columnwidth}
    \includegraphics[width=\mywidth\linewidth]{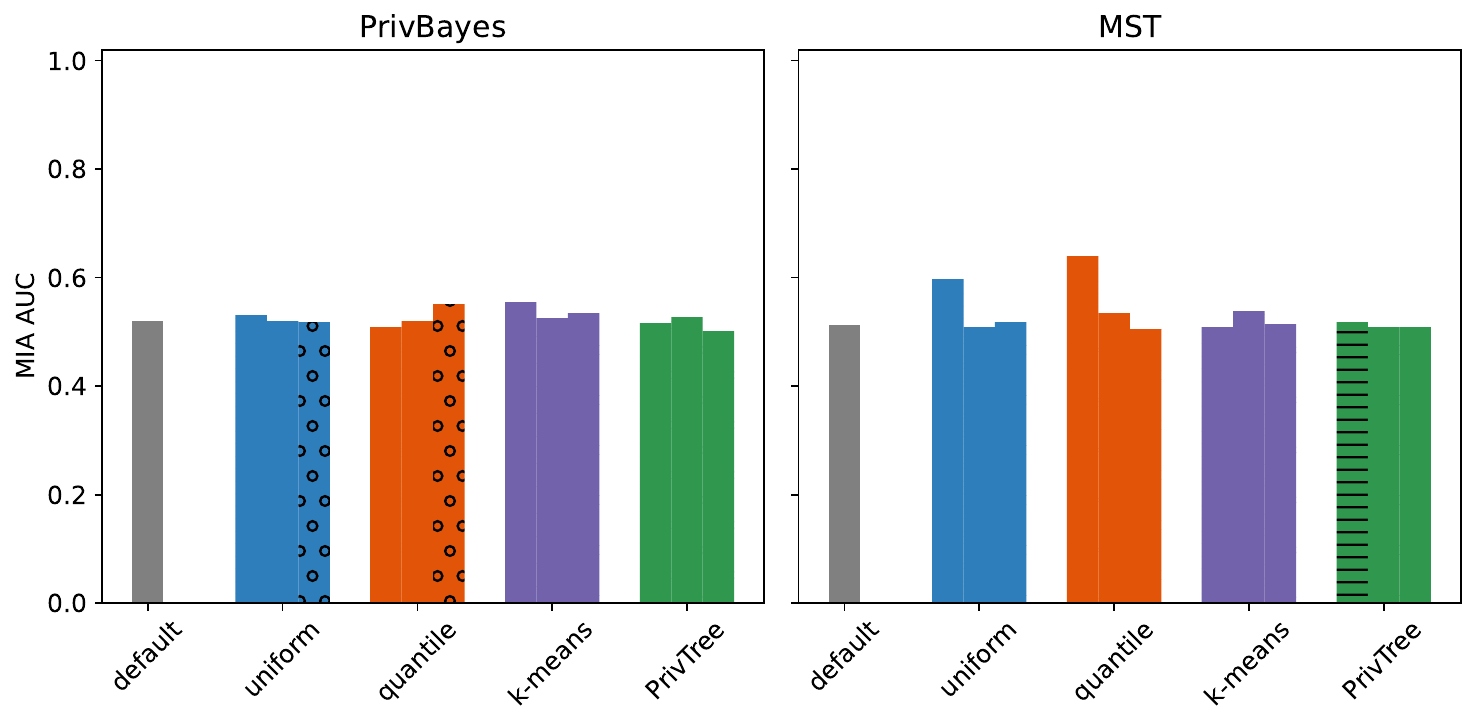}
    \caption{D ($\epsilon=1,000$), G ($\epsilon=1,000$), target \emph{inside} domain}
   	\label{fig:rq4_attack_1000_in}
 	\end{subfigure}
  \caption{Privacy leakage with \emph{provided} and \emph{extracted} domain (w/ and w/o DP) for four DP discretizers~(D) and two DP models~(G) on a target record \emph{outside}/\emph{inside} the domain of the remaining data, GroundHog~\cite{stadler2022synthetic}, Wine dataset (PS1).}
  \reduce
\end{figure*}

\descr{Privacy (PS1).}
\cref{fig:rq4_attack} illustrates the impact of the domain on a popular membership inference attack (MIA), GroundHog~\cite{stadler2022synthetic}, across different discretizers and generative models.\footnote{\chgTag{C7}{We repeat the experiment using Querybased~\cite{houssiau2022tapas} and observe very similar results; due to space limitation, visualizations are provided in~\cref{fig:rq4_attack_q} in Appendix~\ref{app:ps1}.}}
We train the generative models with $\epsilon=1$, and use $\epsilon=1$ for discretization (split in half between domain extraction and discretization for the third setting).
Notably, the selected target record represents a worst-case scenario on a real dataset, which is permitted under DP assumptions and consistent with prior work~\cite{stadler2022synthetic, annamalai2024what, ganev2025elusive}.
The target contains two columns with values significantly larger than those of the remaining records (289 and 440 vs.~146.5 and 366.5).

\chgTag{C6}{Extracting the domain directly from the input data (bars with horizontal lines), particularly the minimum and maximum of numerical columns, without a privacy mechanism not only breaks end-to-end DP guarantees but also exposes highly informative features to the adversary (either directly obtained by GroundHog~\cite{stadler2022synthetic} or indirectly via queries by Querybased~\cite{houssiau2022tapas}).
As a result, the adversary achieves near-perfect performance in all cases, deeming the synthetic data non-private regardless of the discretizer or generative model used.}
This is equivalent to the default strategy (grey bars), which extracts the domain directly and applies \uniform discretization.

By contrast, adopting methods that respect the end-to-end DP pipeline, i.e., either by assuming the domain is provided (bars with crosses) or extracting it with DP (bars with circles), substantially reduces the adversary's success rate.
In all cases, with a single exception (\kmeans with a provided domain), the attack success rate drops to the level of random guessing.
This has several implications.

First, extracting the domain in a DP-compliant way is sufficient to protect against \chgTag{C7}{the specific adversaries used in GroundHog and Querybased attacks}.
Furthermore, the adversaries' success remains low even in settings that could be considered non-private, e.g., (discretizer, generator)-$\epsilon$ values of (100, 100), as shown in~\cref{fig:rq4_attack_100} (again, which the exception of \kmeans).
Similar results are observed for (1, 100), (1, 100), and even (1, 1,000), though not shown due to space constraints.
\chgTag{C6}{The attack only becomes more effective at higher discretizer values,} like (1,000, 1) and (1,000, 1,000) (see~\cref{fig:rq4_attack_1000}); also, recall that it achieves nearly 100\% success when the domain is directly extracted from the data.
This suggests that the effectiveness of \chgTag{C6}{the GroundHog/Querybased attacks} may stem primarily from domain leakage, rather than intrinsic vulnerabilities in the generative model itself.\footnote{These findings are specific to GroundHog~\cite{stadler2022synthetic} \chgTag{C7}{and Querybased~\cite{houssiau2022tapas}, which are among the state-of-the-art MIAs on synthetic data}; domain extraction's impact on attacks tailored to PrivBayes, MST, or other MIAs remains unclear and left to future work.}
To further validate this, we conduct an additional experiment on a target record located farther away from the others but still within their domain (\chgTag{C7}{see~\cref{fig:rq4_attack_1000_in} and~\cref{fig:rq4_attack_1000_in_q} in Appendix~\ref{app:ps1}}) and observe that the attack's success remains near random across all settings.

Second, adopting a DP domain extraction strategy can help address privacy vulnerabilities identified by prior research~\cite{annamalai2024what, ganev2025elusive} in popular model implementations/libraries~\cite{ping2017datasynthesizer, qian2023synthcity} that extract the domain directly from the input data.
In other words, incorporating such techniques can make DP generative model implementations more robust and better align them with end-to-end DP guarantees.

Finally, while extracting the domain in a DP way slightly reduces the adversary's success rate compared to using a provided domain (for $\epsilon \leq 100$), this may come at the cost of utility as already shown in~\cref{fig:rq4_domain}.
Practitioners should therefore prefer using a trusted, externally provided domain (e.g., codebooks for census data) when available, rather than spending additional privacy budget to infer it.
However, further research is needed to explore these tradeoffs and to develop more effective methods for DP domain extraction.

\descr{Take-Aways.}
Although extracting the data domain with DP reduces the utility of synthetic data by 2.5\%–8\% on average, this significantly reduces the success rate of popular MIAs to the level of random guessing, even when applied to worst-case targets.

\section{Related Work}
\label{sec:work}

\descr{DP Generative Models.}
There are several approaches for generating DP tabular synthetic data, including copulas~\cite{li2014differentially, gambs2021growing}, graphical models~\cite{zhang2017privbayes, mckenna2021winning, cai2021data, mahiou2022dpart}, and deep learning models like VAEs~\cite{acs2018differentially, abay2019privacy}, GANs~\cite{xie2018differentially, zhang2018differentially, jordon2018pate, tantipongpipat2021differentially, long2021gpate, xu2023synthetic}, and Diffusion models~\cite{truda2023generating}.
Other approaches include workload/query-based methods~\cite{vietri2020new, aydore2021differentially, liu2021iterative, mckenna2022aim, vietri2022private} and other miscellaneous models~\cite{zhang2021privsyn, ge2021kamino, liu2023generating, vero2024cuts}.
Our study focuses on models from the select-measure-generate paradigm, as they have been proven to consistently perform among the best~\cite{tao2021benchmarking, mckenna2022aim}, particularly for narrow datasets and simple tasks~\cite{ganev2024graphical}.

\descr{Discretization and DP Generative Models.}
As mentioned, the DP models we experiment with expect discretized data as input.
PrivBayes~\cite{zhang2017privbayes} uses \uniform with 8 or 16 bins, extracting the domain directly from the training data.
MST~\cite{mckenna2021winning} relies on 32 equal-width bins, with the domain manually extracted from the training data.\footnote{Please see \cite{pgmissue}. The original MST-NIST implementation~\cite{mckenna2021winning} does not handle discretization as the competition only involved public data and already discretized training data from the same domain~\cite{nist2018differential}, whereas this is not true for the generic MST model.}
RAP~\cite{aydore2021differentially} and GEM~\cite{liu2021iterative} assume discrete data without any discussion but evaluate their models on the data provided by FEM~\cite{vietri2020new}, which discretizes continuous data to 100 bins (before binarizing them).
AIM~\cite{mckenna2022aim} takes a similar approach to MST -- it uses \uniform with 32 bins, assuming the domain is provided.

While it may be well-known in the DP community that preprocessing should always be DP, evidence from the literature indicates that, in practice, this is often ignored and hardly acknowledged~\cite{zhang2017privbayes, ping2017datasynthesizer, vietri2020new, mckenna2021winning, mckenna2022aim, mahiou2022dpart, qian2023synthcity, du2024towards}.
A notable exception is \citet{mckenna2022aim} explicitly noting that manually extracting the domain from the training data may be problematic.\footnote{In Appendix A, \citet{mckenna2022aim} state: {\em \guillemotleft Thus, we ``cheat'' and look at the active domain to automatically derive a domain file from the dataset.\guillemotright} Moreover, in their submission to the NIST competition~\cite{mckenna2021winning}, they note that unique categories should be externally provided -- see~\cite{mcnist} for more details.} %
Also, the dpart DP library~\cite{mahiou2022dpart} issues a {\em PrivacyLeakWarning}, while the OpenDP library~\cite{opendp2021smartnoise} supports DP domain extraction for numerical columns.

In terms of benchmarks,~\citet{tao2021benchmarking} claim that \privtree~\cite{zhang2016privtree}'s discretization outperforms \uniform, alas, omitting any details and experimental evidence.
\citet{du2024towards} run a hyperparameter search on the number of bins using \quantile but only test values between 5 and 20.
By contrast, we compare four different discretization strategies, experiment with a wider number of bins spanning from 5 to 250, and experiment with extracting the domain in a DP way, ensuring that the process is end-to-end DP.

\descr{Discretization and DP Queries.}
As mentioned, the marginal models we experiment with rely on the select-measure-generate paradigm, which can be considered an extension of the select-measure-reconstruct approach for answering linear queries under DP.
In particular, histograms~\cite{nelson2019sok} and data cubes~\cite{ding2011differentially, yaroslavtsev2013accurate, xiao2014dpcube} can be seen as workloads of linear queries (including range and predicate counting queries).
In the most common setting, every data dimension is assumed to be a histogram over categorical or unit-length discrete (consecutive and non-overlapping) bins whose domain is known.

Several papers propose techniques to combine/merge bins or transform the data so that the noise in the measuring step could be distributed more optimally, including clustering~\cite{acs2012differentially, xu2013differentially, zhang2014towards}, hierarchical methods~\cite{hay2010boosting, qardaji2013understanding, zhang2016privtree}, the matrix mechanism~\cite{li2012adaptive, li2015matrix, mckenna2018optimizing}, etc.~\cite{xiao2010differential, li2014data}.
However, the outputs produced by these methods are query answers rather than a trained machine learning model capable of generating high-dimensional data.
Furthermore, they are often limited by the curse of dimensionality, with most being tested on datasets that do not exceed five dimensions~\cite{hay2016principled}.
Finally, these methods do not discuss how one could invert the transformation and how much information would be lost.
In addition to more standard discretization techniques, i.e., \uniform and \quantile (referred to as naive flat methods, which could perform better for datasets with three or more dimensions~\cite{qardaji2013understanding}), we also include one clustering (\kmeans) and one hierarchical (\privtree) approach.

\smallskip\noindent
{\em Relation to Our Own Work:}
A small part of the results from this work also appear in a short paper presented at a proceedings-less workshop~\cite{anonganev2025understanding}.
More precisely, in the workshop paper, we present an experimental evaluation of different strategies for {\em data domain extraction} and their privacy implications, which is closely related to RQ4 in this work.

\section{Conclusion}
This paper studied the impact of Differentially Private (DP) discretization on the utility and privacy of synthetic data produced by DP generative models. %
We emphasized its significance through extensive measurements, evaluating various discretizers and generative models across a wide range of bin counts and privacy budgets.
Our results showed that all DP marginal generative models are highly sensitive to the choice of discretizer and number of bins.
In fact, optimizing these factors can lead to a substantial improvement in utility -- around 30\% on average.

Determining the optimal number of bins is not a straightforward task.
For instance, utility initially improves with more bins but then degrades with an excessive number of bins.
Moreover, training multiple models to identify the best configuration can negatively impact privacy by depleting the privacy budget, besides requiring significant computational resources.
\chgTag{C9}{This motivated us to optimize an existing method for selecting the ideal number of bins, helping bypass the need for extensive model training.}

From a privacy perspective, we demonstrated that the common practice of directly extracting data domain from the training data not only violates end-to-end DP guarantees but also allows adversaries to achieve perfect accuracy on membership inference attacks.
By contrast, using publicly available domain information or extracting data domains in a DP manner provides sufficient protection against the same adversaries, even in worst-case settings.

Overall, our work paves the way toward more trustworthy implementations and deployments of DP generative models and synthetic data, highlighting that not only the model fitting process but the entire pipeline, including pre-processing steps, need to satisfy DP.
We are confident that our work will help data practitioners and algorithm designers alike to better optimize the privacy-utility tradeoffs of DP synthetic tabular data generation pipelines.

\subsection*{Acknowledgments}
We are grateful to the ACM CCS Program Committee for their valuable feedback, which helped us significantly improve our paper.

{\small
\bibliographystyle{plainnat}

\begin{thebibliography}{100}
\providecommand{\natexlab}[1]{#1}
\providecommand{\url}[1]{\texttt{#1}}
\expandafter\ifx\csname urlstyle\endcsname\relax
  \providecommand{\doi}[1]{doi: #1}\else
  \providecommand{\doi}{doi: \begingroup \urlstyle{rm}\Url}\fi

\bibitem[Abay et~al.(2019)Abay, Zhou, Kantarcioglu, Thuraisingham, and
  Sweeney]{abay2019privacy}
Nazmiye~Ceren Abay, Yan Zhou, Murat Kantarcioglu, Bhavani Thuraisingham, and
  Latanya Sweeney.
\newblock {Privacy preserving synthetic data release using deep learning}.
\newblock In \emph{ECML PKDD}, 2019.

\bibitem[Acs et~al.(2012)Acs, Castelluccia, and Chen]{acs2012differentially}
Gergely Acs, Claude Castelluccia, and Rui Chen.
\newblock {Differentially private histogram publishing through lossy
  compression}.
\newblock In \emph{IEEE ICDM}, 2012.

\bibitem[Acs et~al.(2018)Acs, Melis, Castelluccia, and
  De~Cristofaro]{acs2018differentially}
Gergely Acs, Luca Melis, Claude Castelluccia, and Emiliano De~Cristofaro.
\newblock {Differentially private mixture of generative neural networks}.
\newblock \emph{IEEE TKDE}, 2018.

\bibitem[Annamalai et~al.(2024)Annamalai, Ganev, and
  De~Cristofaro]{annamalai2024what}
Meenatchi Sundaram Muthu~Selva Annamalai, Georgi Ganev, and Emiliano
  De~Cristofaro.
\newblock {``What do you want from theory alone?'' Experimenting with Tight
  Auditing of Differentially Private Synthetic Data Generation}.
\newblock In \emph{USENIX Security}, 2024.

\bibitem[Aydore et~al.(2021)Aydore, Brown, Kearns, Kenthapadi, Melis, Roth, and
  Siva]{aydore2021differentially}
Sergul Aydore, William Brown, Michael Kearns, Krishnaram Kenthapadi, Luca
  Melis, Aaron Roth, and Ankit~A Siva.
\newblock {Differentially private query release through adaptive projection}.
\newblock In \emph{ICML}, 2021.

\bibitem[Cai et~al.(2021)Cai, Lei, Wei, and Xiao]{cai2021data}
Kuntai Cai, Xiaoyu Lei, Jianxin Wei, and Xiaokui Xiao.
\newblock {Data synthesis via differentially private markov random fields}.
\newblock \emph{PVLDB}, 2021.

\bibitem[Chaudhuri et~al.(2011)Chaudhuri, Monteleoni, and
  Sarwate]{chaudhuri2011differentially}
Kamalika Chaudhuri, Claire Monteleoni, and Anand~D Sarwate.
\newblock {Differentially private empirical risk minimization}.
\newblock \emph{JMLR}, 2011.

\bibitem[De~Cristofaro(2024)]{cristofaro2024synthetic}
Emiliano De~Cristofaro.
\newblock {Synthetic Data: Methods, Use Cases, and Risks}.
\newblock \emph{IEEE S\&P}, 2024.

\bibitem[Desfontaines(2020)]{desfontaines2020lowering}
Damien Desfontaines.
\newblock \emph{{Lowering the cost of anonymization}}.
\newblock 2020.

\bibitem[Ding et~al.(2011)Ding, Winslett, Han, and Li]{ding2011differentially}
Bolin Ding, Marianne Winslett, Jiawei Han, and Zhenhui Li.
\newblock {Differentially private data cubes: optimizing noise sources and
  consistency}.
\newblock In \emph{SIGMOD}, 2011.

\bibitem[Doane(1976)]{doane1976aesthetic}
David~P. Doane.
\newblock {Aesthetic frequency classifications}.
\newblock \emph{The American Statistician}, 1976.

\bibitem[Du and Li(2024)]{du2024towards}
Yuntao Du and Ninghui Li.
\newblock {Towards Principled Assessment of Tabular Data Synthesis Algorithms}.
\newblock \emph{arXiv:2402.06806}, 2024.

\bibitem[Dua and Graff(2017)]{dua2017data}
Dheeru Dua and Casey Graff.
\newblock {UCI Machine Learning Repository}.
\newblock \url{https://archive.ics.uci.edu/datasets}, 2017.

\bibitem[Dwork et~al.(2006{\natexlab{a}})Dwork, Kenthapadi, McSherry, Mironov,
  and Naor]{dwork2006our}
Cynthia Dwork, Krishnaram Kenthapadi, Frank McSherry, Ilya Mironov, and Moni
  Naor.
\newblock {Our data, ourselves: Privacy via distributed noise generation}.
\newblock In \emph{EuroCrypt}, 2006{\natexlab{a}}.

\bibitem[Dwork et~al.(2006{\natexlab{b}})Dwork, McSherry, Nissim, and
  Smith]{dwork2006calibrating}
Cynthia Dwork, Frank McSherry, Kobbi Nissim, and Adam Smith.
\newblock {Calibrating noise to sensitivity in private data analysis}.
\newblock In \emph{TCC}, 2006{\natexlab{b}}.

\bibitem[Dwork et~al.(2014)Dwork, Roth, et~al.]{dwork2014algorithmic}
Cynthia Dwork, Aaron Roth, et~al.
\newblock {The algorithmic foundations of differential privacy}.
\newblock \emph{Foundations and Trends in Theoretical Computer Science}, 2014.

\bibitem[{FCA}(2024)]{fca2024using}
{FCA}.
\newblock {Using Synthetic Data in Financial Services}.
\newblock
  \url{https://www.fca.org.uk/publication/corporate/report-using-synthetic-data-in-financial-services.pdf},
  2024.

\bibitem[{Forbes}(2022)]{forbes2022synthetic}
{Forbes}.
\newblock {Synthetic data is about to transform artificial intelligence}.
\newblock
  \url{https://www.forbes.com/sites/robtoews/2022/06/12/synthetic-data-is-about-to-transform-artificial-intelligence/},
  2022.

\bibitem[Freedman and Diaconis(1981)]{freedman1981on}
David Freedman and Persi Diaconis.
\newblock {On the histogram as a density estimator:L2 theory}.
\newblock \emph{Probability Theory and Related Fields}, 1981.

\bibitem[Gambs et~al.(2021)Gambs, Ladouceur, Laurent, and
  Roy-Gaumond]{gambs2021growing}
S{\'e}bastien Gambs, Fr{\'e}d{\'e}ric Ladouceur, Antoine Laurent, and Alexandre
  Roy-Gaumond.
\newblock {Growing synthetic data through differentially-private vine copulas}.
\newblock \emph{PETS}, 2021.

\bibitem[Ganev et~al.(2024)Ganev, Xu, and De~Cristofaro]{ganev2024graphical}
Georgi Ganev, Kai Xu, and Emiliano De~Cristofaro.
\newblock {Graphical vs. Deep Generative Models: Measuring the Impact of
  Differentially Private Mechanisms and Budgets on Utility}.
\newblock In \emph{ACM CCS}, 2024.

\bibitem[Ganev et~al.(2025{\natexlab{a}})Ganev, Annamalai, and
  De~Cristofaro]{ganev2025elusive}
Georgi Ganev, Meenatchi Sundaram Muthu~Selva Annamalai, and Emiliano
  De~Cristofaro.
\newblock {The Elusive Pursuit of Reproducing PATE-GAN: Benchmarking, Auditing,
  Debugging}.
\newblock \emph{TMLR}, 2025{\natexlab{a}}.

\bibitem[Ganev et~al.(2025{\natexlab{b}})Ganev, Annamalai, Mahiou, and
  Cristofaro]{anonganev2025understanding}
Georgi Ganev, Meenatchi Sundaram Muthu~Selva Annamalai, Sofiane Mahiou, and
  Emiliano~De Cristofaro.
\newblock {[Tiny] Understanding the Impact of Data Domain Extraction on
  Synthetic Data Privacy}.
\newblock In \emph{ICLR SynthData Workshop}, 2025{\natexlab{b}}.

\bibitem[Ge et~al.(2021)Ge, Mohapatra, He, and Ilyas]{ge2021kamino}
Chang Ge, Shubhankar Mohapatra, Xi~He, and Ihab~F. Ilyas.
\newblock {Kamino: Constraint-Aware Differentially Private Data Synthesis}.
\newblock \emph{PVLDB}, 2021.

\bibitem[Ghosh et~al.(2009)Ghosh, Roughgarden, and
  Sundararajan]{ghosh2009universally}
Arpita Ghosh, Tim Roughgarden, and Mukund Sundararajan.
\newblock {Universally utility-maximizing privacy mechanisms}.
\newblock In \emph{STOC}, 2009.

\bibitem[Golob et~al.(2025)Golob, Pentyala, Maratkhan, and
  De~Cock]{golob2025privacy}
Steven Golob, Sikha Pentyala, Anuar Maratkhan, and Martine De~Cock.
\newblock {Privacy Vulnerabilities in Marginals-based Synthetic Data}.
\newblock In \emph{SaTML}, 2025.

\bibitem[Goodfellow et~al.(2014)Goodfellow, Pouget-Abadie, Mirza, Xu,
  Warde-Farley, Ozair, Courville, and Bengio]{goodfellow2014generative}
Ian Goodfellow, Jean Pouget-Abadie, Mehdi Mirza, Bing Xu, David Warde-Farley,
  Sherjil Ozair, Aaron Courville, and Yoshua Bengio.
\newblock {Generative adversarial nets}.
\newblock \emph{NeurIPS}, 2014.

\bibitem[Hay et~al.(2010)Hay, Rastogi, Miklau, and Suciu]{hay2010boosting}
Michael Hay, Vibhor Rastogi, Gerome Miklau, and Dan Suciu.
\newblock {Boosting the Accuracy of Differentially Private Histograms Through
  Consistency}.
\newblock \emph{PVLDB}, 2010.

\bibitem[Hay et~al.(2016)Hay, Machanavajjhala, Miklau, Chen, and
  Zhang]{hay2016principled}
Michael Hay, Ashwin Machanavajjhala, Gerome Miklau, Yan Chen, and Dan Zhang.
\newblock {Principled evaluation of differentially private algorithms using
  dpbench}.
\newblock In \emph{SIGMOD}, 2016.

\bibitem[Hayes et~al.(2019)Hayes, Melis, Danezis, and
  De~Cristofaro]{hayes2019logan}
Jamie Hayes, Luca Melis, George Danezis, and Emiliano De~Cristofaro.
\newblock {Logan: membership inference attacks against generative models}.
\newblock In \emph{PoPETs}, 2019.

\bibitem[Hod and Canetti(2025)]{hod2025differentially}
Shlomi Hod and Ran Canetti.
\newblock {Differentially Private Release of Israel's National Registry of Live
  Births}.
\newblock In \emph{IEEE S\&P}, 2025.

\bibitem[Holohan et~al.(2019)Holohan, Braghin, Mac~Aonghusa, and
  Levacher]{holohan2019diffprivlib}
Naoise Holohan, Stefano Braghin, P{\'o}l Mac~Aonghusa, and Killian Levacher.
\newblock {Diffprivlib: the IBM differential privacy library}.
\newblock \emph{arXiv:1907.02444}, 2019.
\newblock \url{https://github.com/IBM/differential-privacy-library}.

\bibitem[Houssiau et~al.(2022)Houssiau, Jordon, Cohen, Daniel, Elliott, Geddes,
  Mole, Rangel-Smith, and Szpruch]{houssiau2022tapas}
Florimond Houssiau, James Jordon, Samuel~N Cohen, Owen Daniel, Andrew Elliott,
  James Geddes, Callum Mole, Camila Rangel-Smith, and Lukasz Szpruch.
\newblock {Tapas: a toolbox for adversarial privacy auditing of synthetic
  data}.
\newblock \emph{NeurIPS Workshop on Synthetic Data for Empowering ML Research},
  2022.

\bibitem[Hu et~al.(2024)Hu, Wu, Li, Long, Garrido, Ge, Ding, Forsyth, Li, and
  Song]{hu2024sok}
Yuzheng Hu, Fan Wu, Qinbin Li, Yunhui Long, Gonzalo~Munilla Garrido, Chang Ge,
  Bolin Ding, David Forsyth, Bo~Li, and Dawn Song.
\newblock {Sok: Privacy-preserving data synthesis}.
\newblock In \emph{IEEE S\&P}, 2024.

\bibitem[{IBM}(2024)]{ibm2024creating}
{IBM}.
\newblock {Creating Synthetic data}.
\newblock
  \url{https://www.ibm.com/docs/en/watsonx/saas?topic=data-creating-synthetic},
  2024.

\bibitem[Jordon et~al.(2018)Jordon, Yoon, and Van Der~Schaar]{jordon2018pate}
James Jordon, Jinsung Yoon, and Mihaela Van Der~Schaar.
\newblock {PATE-GAN: Generating synthetic data with differential privacy
  guarantees}.
\newblock In \emph{ICLR}, 2018.

\bibitem[Jordon et~al.(2022)Jordon, Szpruch, Houssiau, Bottarelli, Cherubin,
  Maple, Cohen, and Weller]{jordon2022synthetic}
James Jordon, Lukasz Szpruch, Florimond Houssiau, Mirko Bottarelli, Giovanni
  Cherubin, Carsten Maple, Samuel~N Cohen, and Adrian Weller.
\newblock {Synthetic Data--what, why and how?}
\newblock \emph{arXiv:2205.03257}, 2022.

\bibitem[Koskela and Kulkarni(2024)]{koskela2024practical}
Antti Koskela and Tejas~D Kulkarni.
\newblock {Practical differentially private hyperparameter tuning with
  subsampling}.
\newblock \emph{NeurIPS}, 2024.

\bibitem[Li and Miklau(2012)]{li2012adaptive}
Chao Li and Gerome Miklau.
\newblock {An Adaptive Mechanism for Accurate Query Answering under
  Differential Privacy}.
\newblock \emph{PVLDB}, 2012.

\bibitem[Li et~al.(2014{\natexlab{a}})Li, Hay, Miklau, and Wang]{li2014data}
Chao Li, Michael Hay, Gerome Miklau, and Yue Wang.
\newblock {A Data-and Workload-Aware Algorithm for Range Queries Under
  Differential Privacy}.
\newblock \emph{PVLDB}, 2014{\natexlab{a}}.

\bibitem[Li et~al.(2015)Li, Miklau, Hay, McGregor, and Rastogi]{li2015matrix}
Chao Li, Gerome Miklau, Michael Hay, Andrew McGregor, and Vibhor Rastogi.
\newblock {The matrix mechanism: optimizing linear counting queries under
  differential privacy}.
\newblock \emph{VLDBJ}, 2015.

\bibitem[Li et~al.(2014{\natexlab{b}})Li, Xiong, and
  Jiang]{li2014differentially}
Haoran Li, Li~Xiong, and Xiaoqian Jiang.
\newblock {Differentially private synthesization of multi-dimensional data
  using copula functions}.
\newblock In \emph{EDBT}, 2014{\natexlab{b}}.

\bibitem[Liu et~al.(2021)Liu, Vietri, and Wu]{liu2021iterative}
Terrance Liu, Giuseppe Vietri, and Steven~Z Wu.
\newblock {Iterative methods for private synthetic data: Unifying framework and
  new methods}.
\newblock \emph{NeurIPS}, 2021.

\bibitem[Liu et~al.(2023)Liu, Tang, Vietri, and Wu]{liu2023generating}
Terrance Liu, Jingwu Tang, Giuseppe Vietri, and Steven Wu.
\newblock {Generating private synthetic data with genetic algorithms}.
\newblock In \emph{ICML}, 2023.

\bibitem[Long et~al.(2021)Long, Wang, Yang, Kailkhura, Zhang, Gunter, and
  Li]{long2021gpate}
Yunhui Long, Boxin Wang, Zhuolin Yang, Bhavya Kailkhura, Aston Zhang, Carl~A.
  Gunter, and Bo~Li.
\newblock {G-{PATE}: Scalable Differentially Private Data Generator via Private
  Aggregation of Teacher Discriminators}.
\newblock In \emph{NeurIPS}, 2021.

\bibitem[Mahiou et~al.(2022)Mahiou, Xu, and Ganev]{mahiou2022dpart}
Sofiane Mahiou, Kai Xu, and Georgi Ganev.
\newblock {dpart: Differentially Private Autoregressive Tabular, a General
  Framework for Synthetic Data Generation}.
\newblock \emph{TPDP}, 2022.

\bibitem[McKenna(2019)]{mcnist}
Ryan McKenna.
\newblock
  \url{https://github.com/usnistgov/PrivacyEngCollabSpace/tree/master/tools/de-identification/Differential-Privacy-Synthetic-Data-Challenge-Algorithms/rmckenna},
  2019.

\bibitem[McKenna(2022)]{pgmissue}
Ryan McKenna.
\newblock \url{https://github.com/ryan112358/private-pgm/issues/7}, 2022.

\bibitem[McKenna and Liu(2022)]{mckenna2022simple}
Ryan McKenna and Terrance Liu.
\newblock {A simple recipe for private synthetic data generation}.
\newblock DifferentialPrivacy.org, 2022.
\newblock \url{https://differentialprivacy.org/synth-data-1/}.

\bibitem[McKenna et~al.(2018)McKenna, Miklau, Hay, and
  Machanavajjhala]{mckenna2018optimizing}
Ryan McKenna, Gerome Miklau, Michael Hay, and Ashwin Machanavajjhala.
\newblock {Optimizing error of high-dimensional statistical queries under
  differential privacy}.
\newblock \emph{arXiv preprint arXiv:1808.03537}, 2018.

\bibitem[McKenna et~al.(2019)McKenna, Sheldon, and
  Miklau]{mckenna2019graphical}
Ryan McKenna, Daniel Sheldon, and Gerome Miklau.
\newblock {Graphical-model based estimation and inference for differential
  privacy}.
\newblock In \emph{ICML}, 2019.

\bibitem[McKenna et~al.(2021)McKenna, Miklau, and Sheldon]{mckenna2021winning}
Ryan McKenna, Gerome Miklau, and Daniel Sheldon.
\newblock {Winning the NIST Contest: A scalable and general approach to
  differentially private synthetic data}.
\newblock \emph{JPC}, 2021.

\bibitem[McKenna et~al.(2022)McKenna, Mullins, Sheldon, and
  Miklau]{mckenna2022aim}
Ryan McKenna, Brett Mullins, Daniel Sheldon, and Gerome Miklau.
\newblock {Aim: An adaptive and iterative mechanism for differentially private
  synthetic data}.
\newblock \emph{PVLDB}, 2022.

\bibitem[McSherry and Talwar(2007)]{mcsherry2007mechanism}
Frank McSherry and Kunal Talwar.
\newblock {Mechanism design via differential privacy}.
\newblock In \emph{FOCS}, 2007.

\bibitem[Meeus et~al.(2023)Meeus, Guepin, Cretu, and
  de~Montjoye]{meeus2023achilles}
Matthieu Meeus, Florent Guepin, Ana-Maria Cretu, and Yves-Alexandre
  de~Montjoye.
\newblock {Achilles' Heels: vulnerable record identification in synthetic data
  publishing}.
\newblock \emph{arXiv:2306.10308}, 2023.

\bibitem[Microsoft(2022)]{microsoft2022iom}
Microsoft.
\newblock {IOM and Microsoft release first-ever differentially private
  synthetic dataset to counter human trafficking}.
\newblock
  \url{https://www.microsoft.com/en-us/research/blog/iom-and-microsoft-release-first-ever-differentially-private-synthetic-dataset-to-counter-human-trafficking/},
  2022.

\bibitem[{NASEM}(2020)]{nasem2020census}
{NASEM}.
\newblock \emph{{2020 Census Data Products: Data Needs and Privacy
  Considerations: Proceedings of a Workshop}}.
\newblock The National Academies Press, 2020.

\bibitem[Nelson and Reuben(2019)]{nelson2019sok}
Boel Nelson and Jenni Reuben.
\newblock {Sok: Chasing accuracy and privacy, and catching both in
  differentially private histogram publication}.
\newblock \emph{arXiv:1910.14028}, 2019.

\bibitem[Nikolov et~al.(2013)Nikolov, Talwar, and Zhang]{nikolov2013geometry}
Aleksandar Nikolov, Kunal Talwar, and Li~Zhang.
\newblock {The geometry of differential privacy: the sparse and approximate
  cases}.
\newblock In \emph{ACM STOC}, 2013.

\bibitem[{NIST}(2018)]{nist2018differential}
{NIST}.
\newblock {2018 Differential privacy synthetic data challenge}.
\newblock
  \url{https://www.nist.gov/ctl/pscr/open-innovation-prize-challenges/past-prize-challenges/2018-differential-privacy-synthetic},
  2018.

\bibitem[{NIST}(2020)]{nist2020differential}
{NIST}.
\newblock {2020 Differential privacy temporal map challenge}.
\newblock
  \url{https://www.nist.gov/ctl/pscr/open-innovation-prize-challenges/past-prize-challenges/2020-differential-privacy-temporal},
  2020.

\bibitem[Nowok et~al.(2016)Nowok, Raab, and Dibben]{nowok06synthpop}
Beata Nowok, Gillian~M. Raab, and Chris Dibben.
\newblock {synthpop: Bespoke Creation of Synthetic Data in R}.
\newblock \emph{Journal of Statistical Software}, 2016.

\bibitem[{OECD}(2023)]{oecd2023emerging}
{OECD}.
\newblock {Emerging privacy-enhancing technologies}.
\newblock \url{https://www.oecd-ilibrary.org/content/paper/bf121be4-en}, 2023.

\bibitem[{ONS}(2023)]{ons2023synthesising}
{ONS}.
\newblock {Synthesising the linked 2011 Census and deaths dataset while
  preserving its confidentiality}.
\newblock
  \url{https://datasciencecampus.ons.gov.uk/synthesising-the-linked-2011-census-and-deaths-dataset-while-preserving-its-confidentiality/},
  2023.

\bibitem[{OpenDP}(2021)]{opendp2021smartnoise}
{OpenDP}.
\newblock {SmartNoise SDK: Tools for Differential Privacy on Tabular Data}.
\newblock \url{https://github.com/opendp/smartnoise-sdk}, 2021.

\bibitem[Papernot and Steinke(2022)]{papernot2022hyperparameter}
Nicolas Papernot and Thomas Steinke.
\newblock {Hyperparameter Tuning with Renyi Differential Privacy}.
\newblock In \emph{ICLR}, 2022.

\bibitem[Patki et~al.(2016)Patki, Wedge, and
  Veeramachaneni]{patki2016synthetic}
Neha Patki, Roy Wedge, and Kalyan Veeramachaneni.
\newblock {The synthetic data vault}.
\newblock In \emph{IEEE DSAA}, 2016.

\bibitem[Ping et~al.(2017)Ping, Stoyanovich, and Howe]{ping2017datasynthesizer}
Haoyue Ping, Julia Stoyanovich, and Bill Howe.
\newblock {DataSynthesizer: Privacy-Preserving Synthetic Datasets}.
\newblock In \emph{SSDBM}, 2017.

\bibitem[Qardaji et~al.(2013)Qardaji, Yang, and Li]{qardaji2013understanding}
Wahbeh Qardaji, Weining Yang, and Ninghui Li.
\newblock {Understanding hierarchical methods for differentially private
  histograms}.
\newblock \emph{PVLDB}, 2013.

\bibitem[Qian et~al.(2023)Qian, Davis, and Van Der~Schaar]{qian2023synthcity}
Zhaozhi Qian, Rob Davis, and Mihaela Van Der~Schaar.
\newblock {Synthcity: a benchmark framework for diverse use cases of tabular
  synthetic data}.
\newblock In \emph{NeurIPS Datasets and Benchmarks Track}, 2023.

\bibitem[{Royal Society}(2023)]{rs2023privacy}
{Royal Society}.
\newblock {From privacy to partnership: the role of PETs in data governance and
  collaborative analysis}.
\newblock
  \url{https://royalsociety.org/-/media/policy/projects/privacy-enhancing-technologies/From-Privacy-to-Partnership.pdf},
  2023.

\bibitem[{SAS}(2024)]{sas2024data}
{SAS}.
\newblock {SAS Data Maker}.
\newblock \url{https://www.sas.com/en_gb/software/data-maker.html}, 2024.

\bibitem[Scott(1979)]{scott1979on}
David~W. Scott.
\newblock {On optimal and data-based histograms}.
\newblock \emph{Biometrika}, 1979.

\bibitem[Shimazaki and Shinomoto(2006)]{shimazaki2006recipe}
Hideaki Shimazaki and Shigeru Shinomoto.
\newblock {A recipe for optimizing a time-histogram}.
\newblock In \emph{NIPS}, 2006.

\bibitem[Shokri et~al.(2017)Shokri, Stronati, Song, and
  Shmatikov]{shokri2017membership}
Reza Shokri, Marco Stronati, Congzheng Song, and Vitaly Shmatikov.
\newblock {Membership Inference Attacks against Machine Learning Models}.
\newblock In \emph{IEEE S\&P}, 2017.

\bibitem[Smith(2011)]{smith2011privacy}
Adam Smith.
\newblock {Privacy-preserving statistical estimation with optimal convergence
  rates}.
\newblock In \emph{STOC}, 2011.

\bibitem[Stadler et~al.(2022)Stadler, Oprisanu, and
  Troncoso]{stadler2022synthetic}
Theresa Stadler, Bristena Oprisanu, and Carmela Troncoso.
\newblock {Synthetic Data -- Anonymization Groundhog Day}.
\newblock In \emph{USENIX Security}, 2022.

\bibitem[Sturges(1926)]{sturges1926choice}
Herbert~A Sturges.
\newblock {The choice of a class interval}.
\newblock \emph{JASA}, 1926.

\bibitem[Su et~al.(2016)Su, Cao, Li, Bertino, and Jin]{su2016differentially}
Dong Su, Jianneng Cao, Ninghui Li, Elisa Bertino, and Hongxia Jin.
\newblock {Differentially private k-means clustering}.
\newblock In \emph{CODASPY}, 2016.

\bibitem[Tantipongpipat et~al.(2021)Tantipongpipat, Waites, Boob, Siva, and
  Cummings]{tantipongpipat2021differentially}
Uthaipon~Tao Tantipongpipat, Chris Waites, Digvijay Boob, Amaresh~Ankit Siva,
  and Rachel Cummings.
\newblock {Differentially private synthetic mixed-type data generation for
  unsupervised learning}.
\newblock \emph{IDT}, 2021.

\bibitem[Tao et~al.(2022)Tao, McKenna, Hay, Machanavajjhala, and
  Miklau]{tao2021benchmarking}
Yuchao Tao, Ryan McKenna, Michael Hay, Ashwin Machanavajjhala, and Gerome
  Miklau.
\newblock {Benchmarking differentially private synthetic data generation
  algorithms}.
\newblock \emph{PPAI}, 2022.

\bibitem[{TechCrunch}(2022)]{techcrunch2022the}
{TechCrunch}.
\newblock {The market for synthetic data is bigger than you think}.
\newblock
  \url{https://techcrunch.com/2022/05/10/the-market-for-synthetic-data-is-bigger-than-you-think/},
  2022.

\bibitem[Truda(2023)]{truda2023generating}
Gianluca Truda.
\newblock {Generating tabular datasets under differential privacy}.
\newblock \emph{arXiv:2308.14784}, 2023.

\bibitem[{UK ICO}(2023{\natexlab{a}})]{ico2023privacy}
{UK ICO}.
\newblock {Privacy-enhancing technologies (PETs)}.
\newblock
  \url{https://ico.org.uk/media/for-organisations/uk-gdpr-guidance-and-resources/data-sharing/privacy-enhancing-technologies-1-0.pdf},
  2023{\natexlab{a}}.

\bibitem[{UK ICO}(2023{\natexlab{b}})]{ico2023synthetic}
{UK ICO}.
\newblock {Synthetic data to test the effectiveness of a vulnerable person's
  detection system in financial services}.
\newblock
  \url{https://ico.org.uk/for-organisations/uk-gdpr-guidance-and-resources/data-sharing/privacy-enhancing-technologies/case-studies/synthetic-data-to-test-the-effectiveness-of-a-vulnerable-persons-detection-system-in-financial-services/},
  2023{\natexlab{b}}.

\bibitem[{UN}(2023)]{un2023guide}
{UN}.
\newblock {The United Nations Guide on privacy-enhancing technologies for
  official statistics}.
\newblock
  \url{https://unstats.un.org/bigdata/task-teams/privacy/guide/2023_UN\%20PET\%20Guide.pdf},
  2023.

\bibitem[Vero et~al.(2024)Vero, Balunovi{\'c}, and Vechev]{vero2024cuts}
Mark Vero, Mislav Balunovi{\'c}, and Martin Vechev.
\newblock {CuTS: Customizable Tabular Synthetic Data Generation}.
\newblock \emph{ICML}, 2024.

\bibitem[Vietri et~al.(2020)Vietri, Tian, Bun, Steinke, and Wu]{vietri2020new}
Giuseppe Vietri, Grace Tian, Mark Bun, Thomas Steinke, and Steven Wu.
\newblock {New oracle-efficient algorithms for private synthetic data release}.
\newblock In \emph{ICML}, 2020.

\bibitem[Vietri et~al.(2022)Vietri, Archambeau, Aydore, Brown, Kearns, Roth,
  Siva, Tang, and Wu]{vietri2022private}
Giuseppe Vietri, Cedric Archambeau, Sergul Aydore, William Brown, Michael
  Kearns, Aaron Roth, Ankit Siva, Shuai Tang, and Steven~Z Wu.
\newblock {Private synthetic data for multitask learning and marginal queries}.
\newblock \emph{NeurIPS}, 2022.

\bibitem[Xiao et~al.(2010)Xiao, Wang, and Gehrke]{xiao2010differential}
Xiaokui Xiao, Guozhang Wang, and Johannes Gehrke.
\newblock {Differential privacy via wavelet transforms}.
\newblock \emph{IEEE TKDE}, 2010.

\bibitem[Xiao et~al.(2014)Xiao, Xiong, Fan, and Goryczka]{xiao2014dpcube}
Yonghui Xiao, Li~Xiong, Liyue Fan, and Slawomir Goryczka.
\newblock {DPCube: Differentially private histogram release through
  multidimensional partitioning}.
\newblock \emph{TDP}, 2014.

\bibitem[Xie et~al.(2018)Xie, Lin, Wang, Wang, and Zhou]{xie2018differentially}
Liyang Xie, Kaixiang Lin, Shu Wang, Fei Wang, and Jiayu Zhou.
\newblock {Differentially private generative adversarial network}.
\newblock \emph{arXiv:1802.06739}, 2018.

\bibitem[Xu et~al.(2013)Xu, Zhang, Xiao, Yang, Yu, and
  Winslett]{xu2013differentially}
Jia Xu, Zhenjie Zhang, Xiaokui Xiao, Yin Yang, Ge~Yu, and Marianne Winslett.
\newblock {Differentially private histogram publication}.
\newblock \emph{VLDBJ}, 2013.

\bibitem[Xu et~al.(2023)Xu, Ganev, Joubert, Davison, Van~Acker, and
  Robinson]{xu2023synthetic}
Kai Xu, Georgi Ganev, Emile Joubert, Rees Davison, Olivier Van~Acker, and Luke
  Robinson.
\newblock {Synthetic data generation of many-to-many datasets via random graph
  generation}.
\newblock In \emph{ICLR}, 2023.

\bibitem[Yaroslavtsev et~al.(2013)Yaroslavtsev, Cormode, Procopiuc, and
  Srivastava]{yaroslavtsev2013accurate}
Grigory Yaroslavtsev, Graham Cormode, Cecilia~M Procopiuc, and Divesh
  Srivastava.
\newblock {Accurate and efficient private release of datacubes and contingency
  tables}.
\newblock In \emph{IEEE ICDE}, 2013.

\bibitem[Zhang et~al.(2016)Zhang, Xiao, and Xie]{zhang2016privtree}
Jun Zhang, Xiaokui Xiao, and Xing Xie.
\newblock {Privtree: A differentially private algorithm for hierarchical
  decompositions}.
\newblock In \emph{SIGMOD}, 2016.

\bibitem[Zhang et~al.(2017)Zhang, Cormode, Procopiuc, Srivastava, and
  Xiao]{zhang2017privbayes}
Jun Zhang, Graham Cormode, Cecilia~M Procopiuc, Divesh Srivastava, and Xiaokui
  Xiao.
\newblock {Privbayes: Private data release via bayesian networks}.
\newblock \emph{ACM TODS}, 2017.

\bibitem[Zhang et~al.(2014)Zhang, Chen, Xu, Meng, and Xie]{zhang2014towards}
Xiaojian Zhang, Rui Chen, Jianliang Xu, Xiaofeng Meng, and Yingtao Xie.
\newblock {Towards accurate histogram publication under differential privacy}.
\newblock In \emph{SDM}, 2014.

\bibitem[Zhang et~al.(2018)Zhang, Ji, and Wang]{zhang2018differentially}
Xinyang Zhang, Shouling Ji, and Ting Wang.
\newblock {Differentially private releasing via deep generative model
  (technical report)}.
\newblock \emph{arXiv:1801.01594}, 2018.

\bibitem[Zhang et~al.(2021)Zhang, Wang, Honorio, Li, Backes, He, Chen, and
  Zhang]{zhang2021privsyn}
Zhikun Zhang, Tianhao Wang, Jean Honorio, Ninghui Li, Michael Backes, Shibo He,
  Jiming Chen, and Yang Zhang.
\newblock {PrivSyn: Differentially Private Data Synthesis}.
\newblock In \emph{USENIX Security}, 2021.

\end{thebibliography}

}

\appendix

\begin{figure}[h!]
	\centering
	\includegraphics[width=0.99\linewidth]{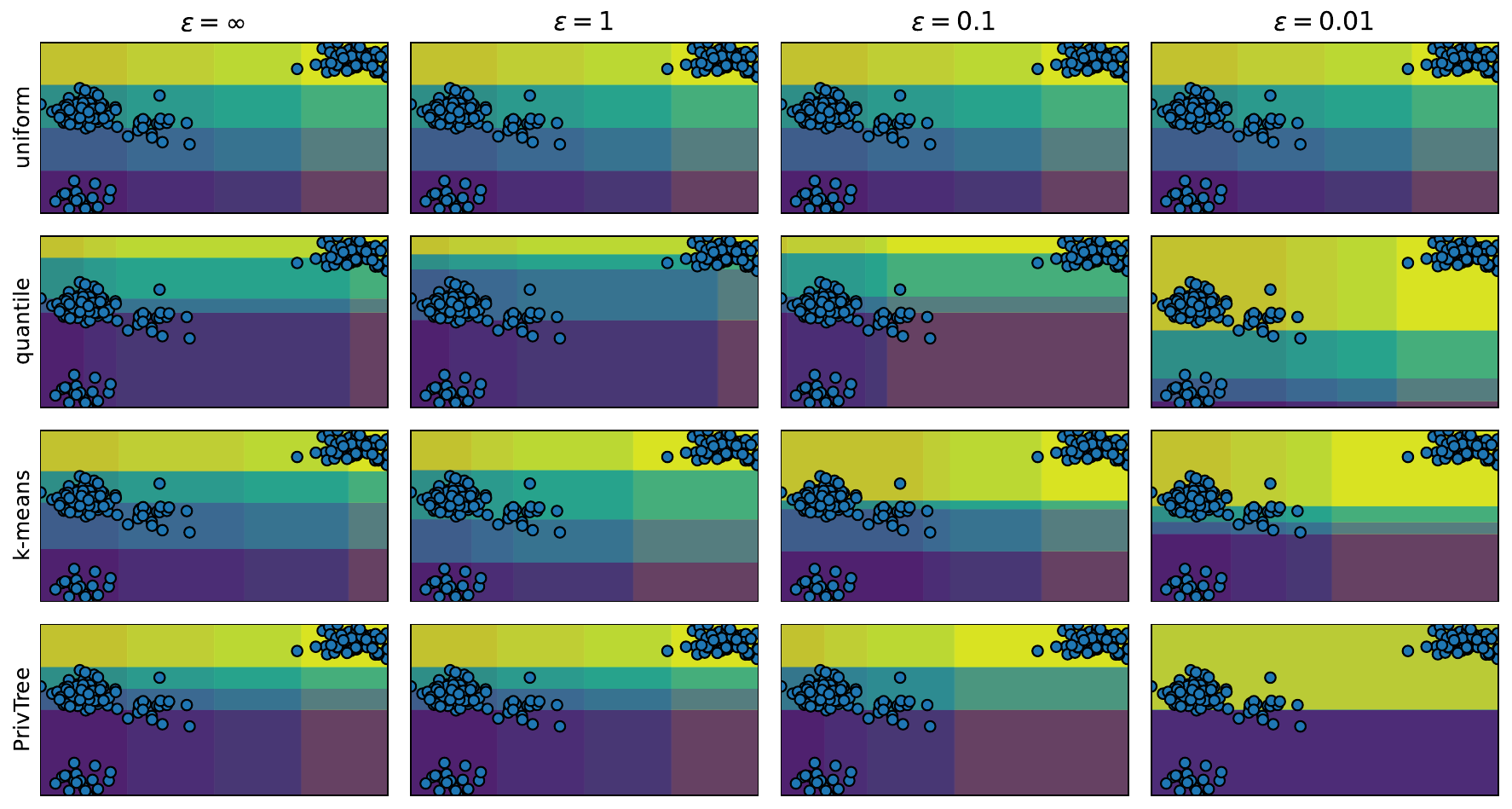}
	\caption{Reproduction of scikit-learn's clustering example with added DP discretization.}
	\label{fig:clustering}
	\reduce
\end{figure}

\begin{figure*}[t!]
	\centering
	\includegraphics[width=0.99\linewidth]{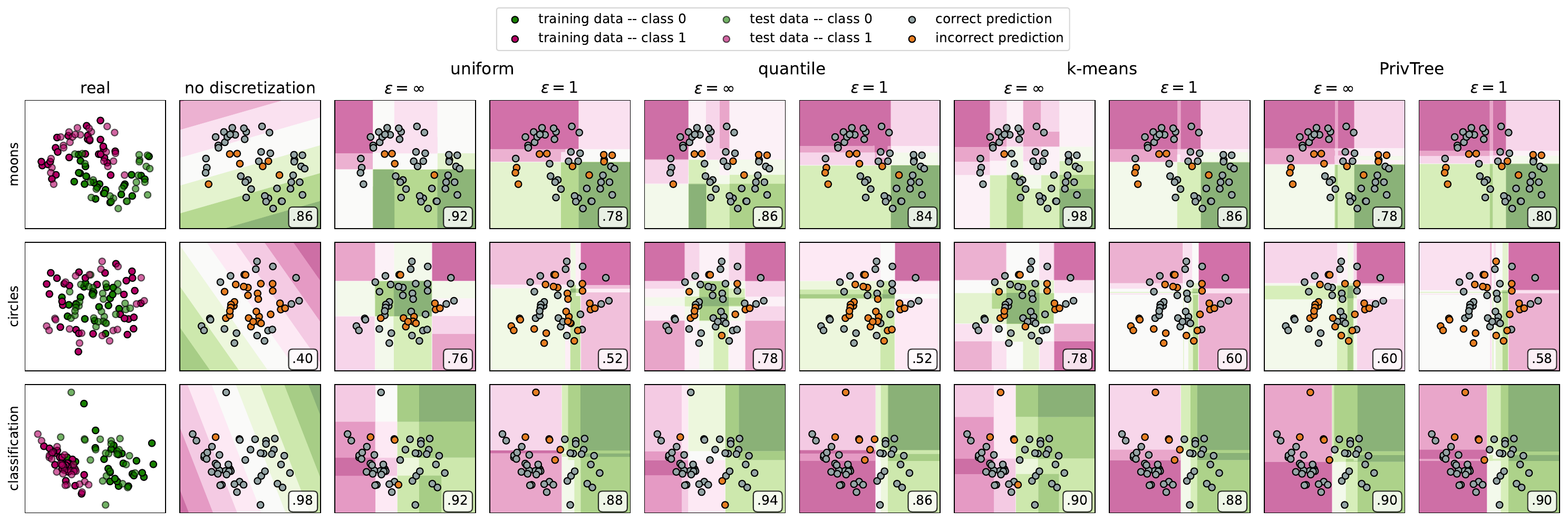}
	\caption{Reproduction of scikit-learn's classification example with added DP discretization.}
	\label{fig:classification}
	\reduce
\end{figure*}

\begin{figure*}[t!]
  \begin{subfigure}[t]{0.99\linewidth}
    \centering
    \includegraphics[width=0.6\linewidth]{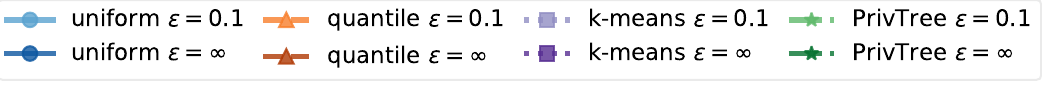}
    \vspace{-0.35cm}
  \end{subfigure}
  \centering
  \begin{subfigure}[t]{0.9\linewidth}
    \includegraphics[width=\mywidth\linewidth]{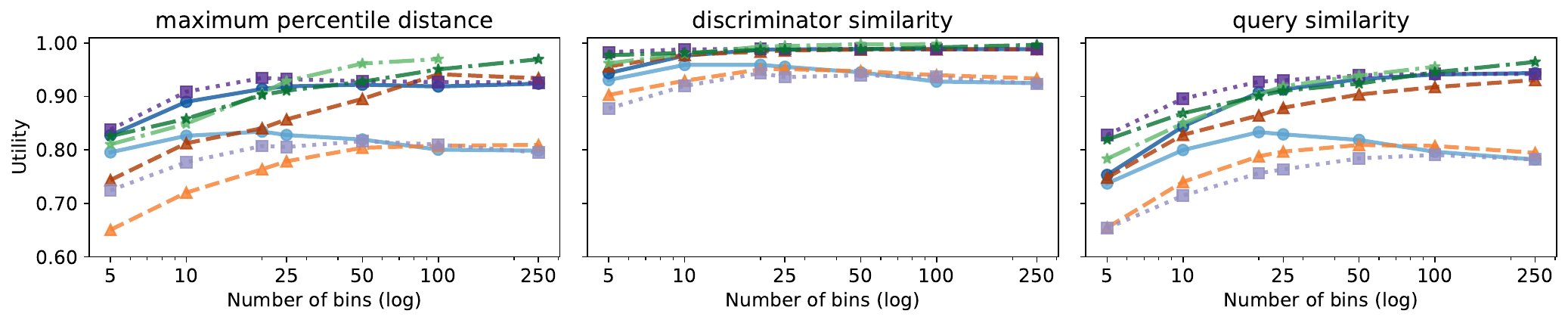}
	\end{subfigure}
  \caption{Utility of DP discretizers w/ baseline DP generative modeling, averaged across five controlled distributions (US2).}
  \label{fig:rq2_bins_1d_dis}
  \reduce
\end{figure*}

\begin{figure*}[t!]
  \begin{subfigure}[t]{0.99\linewidth}
    \centering
    \includegraphics[width=0.9\linewidth]{plots/rq4/legend.pdf}
    \vspace{-0.35cm}
  \end{subfigure}
  \centering
  \begin{subfigure}[t]{0.23\linewidth}
    \includegraphics[width=\mywidth\linewidth]{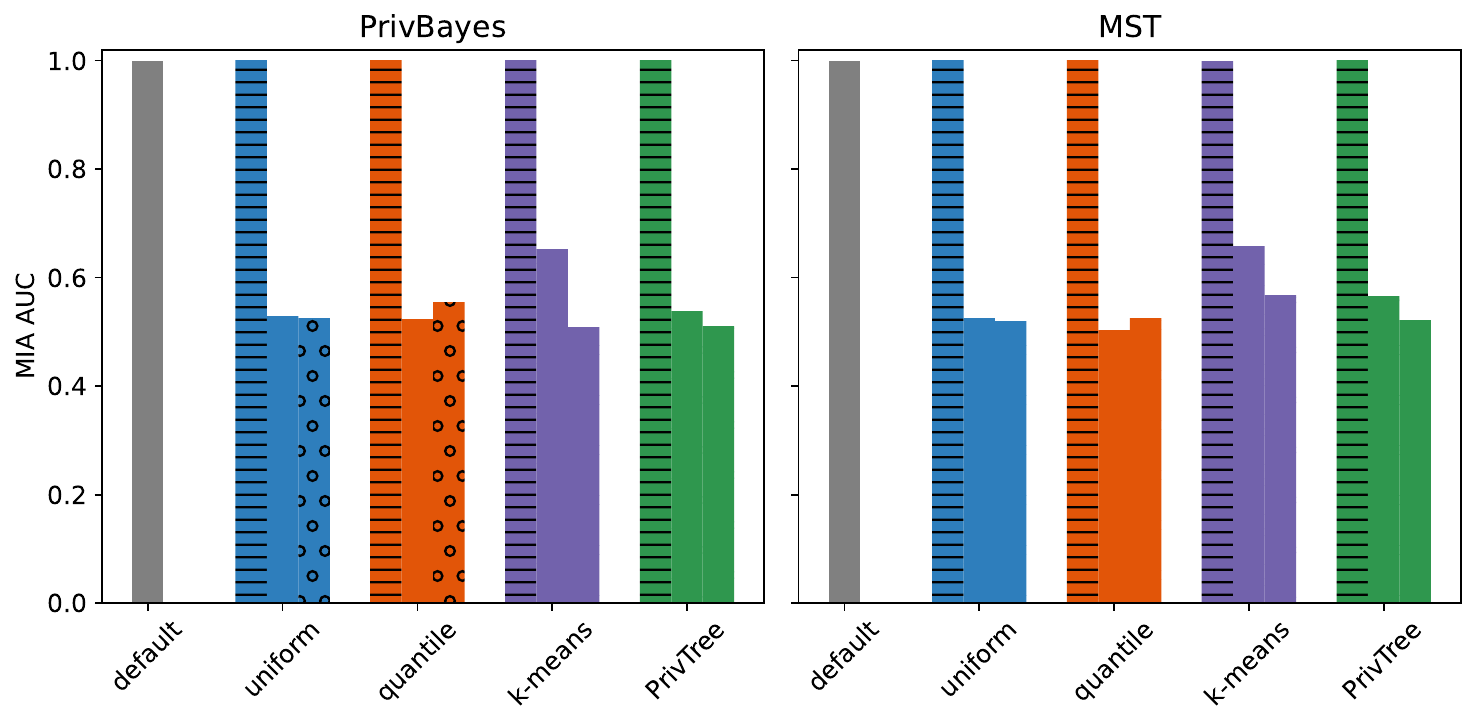}
    \caption{\footnotesize D ($\epsilon=1$), G ($\epsilon=1$), \emph{outside}}
  	\label{fig:rq4_attack_1_q}
	\end{subfigure}
  \begin{subfigure}[t]{0.23\linewidth}
    \includegraphics[width=\mywidth\linewidth]{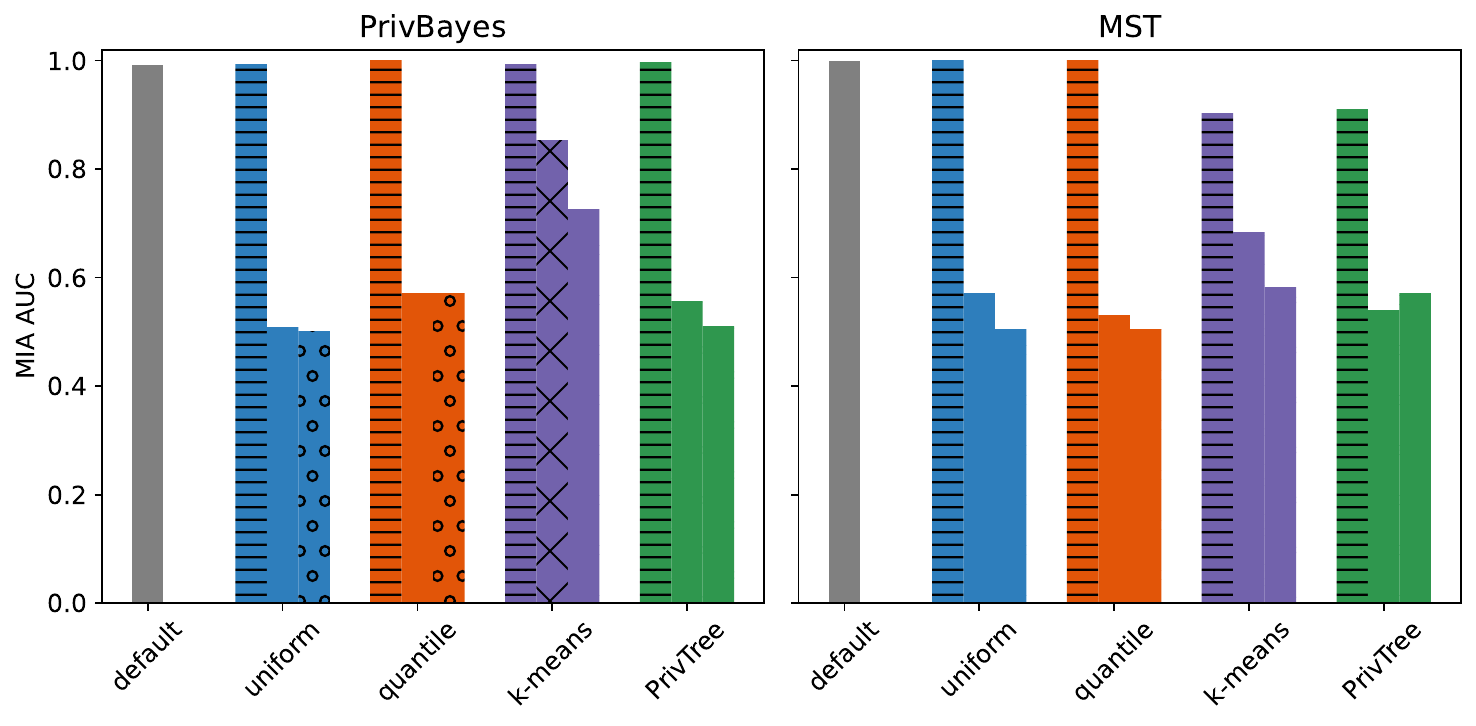}
    \caption{\footnotesize D ($\epsilon=100$), G ($\epsilon=100$), \emph{outside}}
   	\label{fig:rq4_attack_100_q}
 	\end{subfigure}
  \begin{subfigure}[t]{0.23\linewidth}
    \includegraphics[width=\mywidth\linewidth]{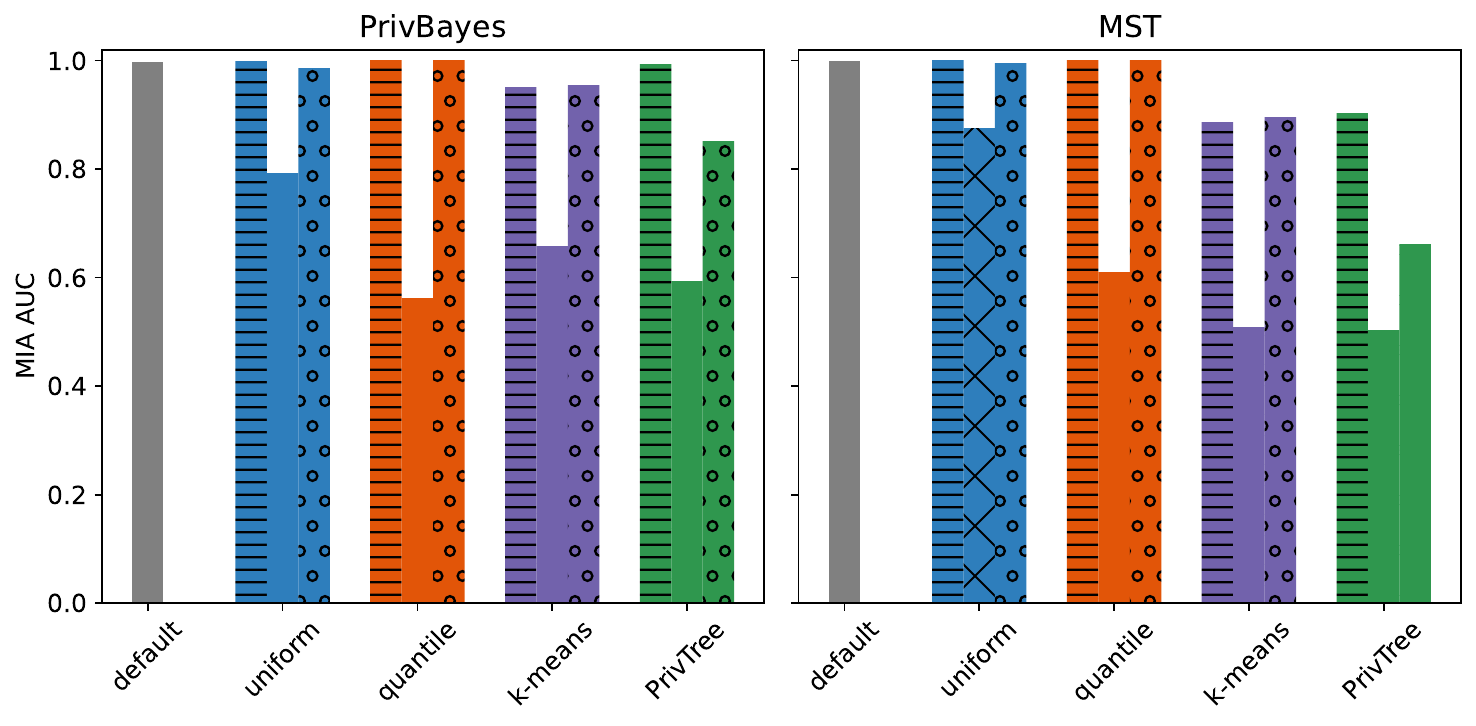}
    \caption{\footnotesize D ($\epsilon=1,000$), G ($\epsilon=1,000$), \emph{outside}}
   	\label{fig:rq4_attack_1000_q}
  \end{subfigure}
  \begin{subfigure}[t]{0.23\linewidth}
    \includegraphics[width=\mywidth\linewidth]{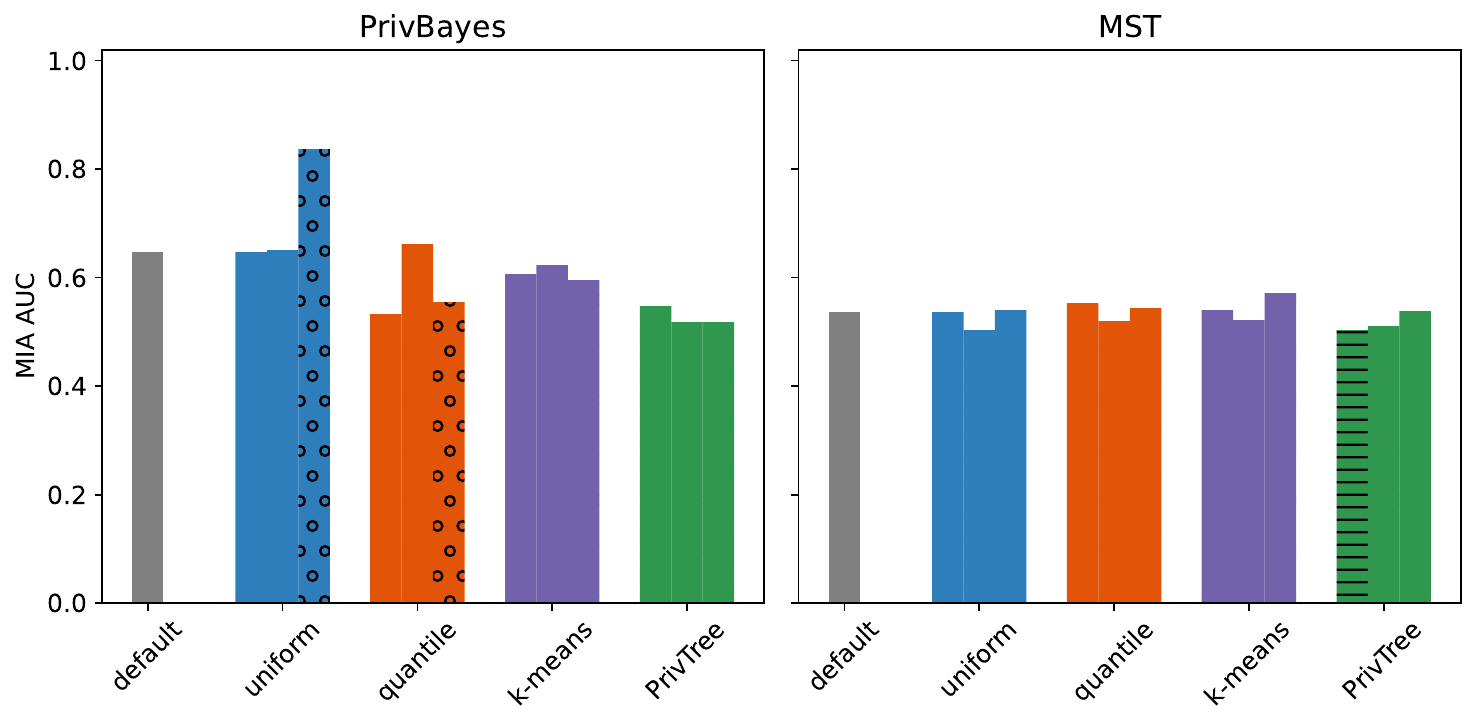}
    \caption{\footnotesize D ($\epsilon=1,000$), G ($\epsilon=1,000$), \emph{inside}}
   	\label{fig:rq4_attack_1000_in_q}
 	\end{subfigure}
  \caption{Privacy leakage with \emph{provided} and \emph{extracted} domain (w/ and w/o DP) for four DP discretizers~(D) and two DP models~(G) on a target record \emph{outside}/\emph{inside} the domain of the remaining data, Querybased~\cite{houssiau2022tapas}, Wine dataset (PS1).}
  \label{fig:rq4_attack_q}
  \reduce
\end{figure*}

\section{Scikit-Learn Examples}
\label{app:exp}
In~\cref{fig:clustering} and~\ref{fig:classification}, we present two examples, visually highlighting the importance of discretization in clustering and classification.
(Inspired by Scikit-Learn, see \url{https://scikit-learn.org/stable/auto_examples/preprocessing/plot_discretization_strategies.html} as well as \url{https://scikit-learn.org/stable/auto_examples/preprocessing/plot_discretization_classification.html}).

In the first example, shown in~\cref{fig:clustering}, we apply DP discretization to a clustering task on a toy dataset.
The figure uses different colors to denote regions where the discretized encoding remains constant.
It illustrates how different discretizers encode the data differently and, except for uniform, how the privacy budget influences the shape, size, and number of resulting bins.
For instance, \quantile does not perform particularly well, as it merges records from different clusters across all budgets.
Unsurprisingly, \kmeans performs consistently well, while \privtree reduces the number of bins as the privacy budget decreases.

The second example, visualized in~\cref{fig:classification}, compares the performance of logistic regression with and without DP discretization on three toy datasets.
The shade of the colors reflects the confidence of the fitted classifiers, highlighting how discretization affects decision boundaries.
As expected, discretization does not help in the nearly linearly separable case (third row).
However, in the other two cases, all discretizers except one, at $\epsilon=\infty$, produce data representations that improve classifier accuracy.
Notably, classifiers trained on discretized data with $\epsilon=1$ outperform the baseline for the circles data (second row).
This improvement is likely due to the limitations of logistic regression; more advanced classifiers, like random forests, would not face these challenges initially.
Nevertheless, this example shows that discretization can enhance modeling in certain scenarios.
Overall, both \quantile and \kmeans discretizers demonstrate strong performance, while \privtree underperforms.

\section{\chgTag{C10}{RQ2: Discretization and DP Generative Models (Additional Experiment)}}
\label{app:us2}
\chg{In~\cref{fig:rq2_bins_1d_dis}, we present disaggregated metrics (namely, maximum percentile distance, discriminator similarity, and query similarity) of the four discretizers across a range of bins and two privacy budgets in US2.
The overall behavior of the discretizers is consistent across the metrics, as discussed in~\cref{subsec:q2}.}

\section{\chgTag{C7}{RQ4: Discretization and DP Domain Extraction (Additional Experiment)}}
\label{app:ps1}
\chg{In~\cref{fig:rq4_attack_q}, we present results from running Querybased~\cite{houssiau2022tapas} in PS1 on the Wine dataset with three data domain extraction strategies and two target records.
The findings closely mirror those obtained with GroundHog~\cite{stadler2022synthetic}, as discussed in~\cref{subsec:q4}.}

\end{document}